\documentclass[11pt]{article}
    \usepackage{caption3}
    \DeclareCaptionOption{parskip}[]{}
    \usepackage[small]{caption}
\usepackage{amsmath,amssymb,verbatim}
\usepackage{natbib}
\usepackage{color}
\usepackage{hyperref}
\usepackage{longtable}
\usepackage{lipsum}
\usepackage{boxedminipage}
\usepackage{fancybox}
\usepackage{bbm}
\usepackage{amsfonts}
\usepackage{makecell}
\usepackage{graphicx}
\usepackage{setspace}

\newcommand{\indic}[1]{1\hspace{-2.1mm}{1}_{\{#1\}}} 
\newtheorem{theorem}{Theorem}

\newtheorem{proposition}[theorem]{Proposition}
\newtheorem{remark}[theorem]{Remark}

\newtheorem{corollary}[theorem]{Corollary}

\def\Q{{\mathbb Q}}        
\def\R{{\mathbb R}}        
\def\P{{\mathbb P}}        
\def\E{{\mathbb E}}        
\def\1{{\mathbf 1}}        
\def\Fil{{\mathbb F}}        

\def\setT{{\mathcal T}}

\def\Cvest{{C}}
\def\Cunvest{{\tilde{C}}}

\def\tl{{\tau^\lambda}}

 \def\lv{L\'{e}vy }

\numberwithin{equation}{section}
\numberwithin{theorem}{section}
\begin{document}
\title{ESO  Valuation with Job Termination Risk and Jumps in Stock Price}
\author{Tim Leung\thanks{Industrial Engineering \& Operations Research Department, Columbia University, New York, NY 10027; email:\,\mbox{leung@ieor.columbia.edu}. Corresponding author. } \and Haohua Wan\thanks{Industrial and Enterprise Systems Engineering
 Department, University of Illinois at Urbana-Champaign, Urbana, IL 61801; email:\,\mbox{hwan3@illinois.edu}.} }
 \maketitle
 \begin{abstract}
 Employee stock  options (ESOs) are American-style call options  that can  be terminated early due to employment shock.   This paper studies an ESO  valuation framework that accounts for  job termination risk  and   jumps in the company stock price.    Under general L\'evy stock price dynamics, we show that a higher job termination risk induces the ESO holder to voluntarily accelerate  exercise, which in turn reduces the   cost to  the company. The  holder's optimal exercise boundary and ESO cost  are determined by solving  an  inhomogeneous partial integro-differential variational inequality (PIDVI).  We  apply  Fourier transform  to simplify the variational inequality and develop  accurate  numerical methods.  Furthermore, when the stock price follows a geometric Brownian motion, we provide closed-form formulas for both the vested and unvested perpetual  ESOs.  Our model is also applied  to   evaluate  the probabilities of understating ESO expenses  and   contract termination.
  \end{abstract}

\noindent {\textbf{Keywords:}\,  employee stock option, job termination, L\'evy processes, Fourier transform 

 \newpage

\section{Introduction}\label{sect-intro}
Employee stock options (ESOs) are an integral part of  executive  compensation in the United States. The primary objective is to align the interests between the executives and  the firm. According to \cite{Frydman2010},  stock options account for 25\% of the total compensation package of CEOs in 2008.  In Table 1, we see that the percentage of  S\&P 500 companies granting ESOs remains above 73\% in 2011, despite a decline from 93\% in 2000.  The cost of these options
can potentially be very burdensome to the shareholders.
In view of this, the Financial Accounting Standards Board (FASB) has passed regulations to require firms to estimate and report the granting cost of ESOs.

 Typically, ESOs are early-exercisable long-dated call options  written on the company stock. To  maintain the incentive effect, the firm usually  imposes a \emph{vesting period} that prohibits the employee from exercising the option.  During the vesting period,  the employee's departure from the firm will result in forfeiture of the option (i.e. it becomes worthless).  After the vesting period, when the employee leaves the firm,   the ESO will expire though the employee can choose to exercise if the option is in the money\footnote{See, for example,  ``A Detailed Overview of Employee Ownership Plan Alternatives" by The National Center for Employee Ownership (http://www.nceo.org). }. Table \ref{Tab_1} summarizes  the average vesting period and average maturity of ESOs granted  by S\&P 500 companies over  2000-2011. As we can see, the vesting period has been consistently close to 2 years, while the average maturity has decreased   from 9-10 years to 8 years over  a decade.

The key to ESO valuation involves modeling the employee's voluntary exercise strategy   as well as  job termination time, especially since the option is typically  terminated prior to the contractual expiration date.  Moreover, the possibility of future employment shock can influence the employee's decision to exercise now or later.  In fact, the FASB guideline\footnote{See Sect. A.16, FASB Statement 123R (revised 2004), available  on http://www.fasb.org/summary/stsum123r.shtml. } also recommends that any reasonable  ESO valuation model  incorporate ``the effects of employees' expected exercise and post-vesting employment termination behavior."

In this paper, we study a valuation framework that  incorporates the common ESO features of vesting period, early exercise and job termination risk, while allowing for different price dynamics with jumps.  Specifically, we model the arrival of the employment shock by an exogenous jump process, and   formulate the American-style ESO as an optimal stopping problem with possible forced exercise prior to expiration date. Our valuation problem is studied under a class of exponential  L\'evy  price processes, rather than limiting to the GBM framework commonly found in the literature (see e.g. \cite{HullWhite04,Cvitanic2008,Carpenter10}). Under different job termination intensity assumptions (constant or stochastic), we analyze  the corresponding free boundary problems in terms of an inhomogeneous partial integro-differential  variational inequality (PIDVI), and discuss  the computational methods to solve for the option value as well as the optimal exercise strategy. Analytically and numerically, we find that with higher job termination risk it is optimal for the holder to voluntarily accelerate ESO exercise.  For risk analysis, we also apply our  numerical schemes to calculate the probability of cost exceedance  and the probability of contract termination under various scenarios.

 In existing literature, there are several major approaches to model the exercise  timing of ESOs. The first approach assigns an  ad hoc  exercise boundary for the ESO holder.  In other words,  the ESO exercise is characterized by the first passage  time of underlying stock price to a pre-specified level, which is independent of the employee's job termination rate or other model parameters. The main advantage of this approach is the availability of closed-form formulas for the ESO value. Additionally, one can also incorporate a random job termination time, as in  \cite{HullWhite04,Cvitanic2008}, but it would not directly affect the voluntary exercise boundary.

  To incorporate the employee's risk  attitude, the \emph{utility maximization} approach derives the employee's exercise  strategy from a portfolio optimization problem  that accounts for the employee's risk aversion and hedging restrictions \citep{Huddart94,Chance04,GrasselliHenderson2008,LeungSircarESO_MF09,Carpenter10}. The utility-maximizing   exercise boundary represents the optimal voluntary exercise timing for  the risk-averse employee. From the firm's perspective, the ESO cost is computed by risk-neutral expectation assuming the option will be exercised at the utility-maximizing boundary or  job termination time, whichever is earlier.

On the other hand, the \emph{intensity-based} approach does not distinguish   voluntary and involuntary exercise, and model the contract termination by a single random time, characterized by the first arrival time of an exogenous jump process.  In the literature,  \cite{JennergenNaslund93} model the exercise time using an exogenous Poisson process, and \cite{CarrLin00} propose an intensity-based model for  ESO valuation where both  job termination and voluntary exercise intensities are functions of the company stock price.  In the theoretical study by \cite{Szimayer2004}, the employee is allowed to optimally select a voluntary exercise time, while the sudden departure  is represented  by an exogenous Cox process.   In practice, the  job termination intensity specification depends on the firm's history and estimation methodology. For empirical studies on the early exercise patterns of ESOs, we refer to \cite{HuddartLang96,Marquardt2002,Bettis}.

\begin{table}
\begin{center}
\begin{tabular}{lcccccccccccccccccccccc}
\hline
Year &   \% of S\&P500 companies  & Avg. vesting period      & Avg. Maturity        \\
 & granting ESOs &  (years) & (years)  \vspace{2pt} \\
\hline
\hline
2000  & 92.98\% & 2.00&9.24  \\
2001  & 94.35\%& 2.22&9.28 \\
2002   & 93.56\%& 2.18&9.53 \\
2003   & 89.59\%& 2.18&10.17 \\
2004   & 88.09\%& 2.03&8.66 \\
2005   & 75.34\%& 2.16&8.61 \\
2006   & 81.30\% & 2.12&7.86 \\
2007   & 77.33\% & 2.18&8.14 \\
2008   & 76.31\% & 2.38&7.35 \\
2009   & 75.30\% & 2.16&8.41 \\
2010   & 72.69\% & 2.32&8.71 \\
2011  & 73.39\% & 2.16 & 8.07\\
\hline\end{tabular}
\end{center}
\small \caption{Summary of ESO compensation during 2000-2011. Source: Thomson Reuters Insider Trading database and Compustat ExecuComp database.}\label{Tab_1}
\end{table}

We consider  the ESO as an American call option with a possible sudden forfeiture  or forced exercise due to job termination. In contrast to many existing ESO models, we study a versatile  valuation framework that is compatible for a wide class of  \lv price processes, including \cite{Merton_jump1976} and \cite{Kou2002} jump diffusions, as well as Variance Gamma (VG) (\cite{MadanCarr1998}) and  CGMY (\cite{MadanCarr2002}) models, in addition to  GBM that is commonly found in the literature including those cited above. This allows us to study the combined effect of jumps in stock price and job termination intensity on the ESO value and optimal exercise boundary. Moreover,  our model provides  the end-user, i.e. the firm,  the flexibility in choosing the appropriate price model (within a general \lv class) for the underlying stock.  Currently, the FASB permits  the  use of the Black-Scholes  formula with the ESO's contractual term replaced by its average life,  as well as a number of variations\footnote{See  Sect. A.25, FASB Statement 123R (revised 2004).}. In this regard,   our paper  offers an alternative valuation approach that accounts for both voluntary exercise and  job termination risk along  with various choices of models for  the  company stock price.

The  ESO valuation problem  can be considered as an extension to the pricing of  American options under \lv processes; see  \cite{Pham1997}, \cite{MadanHirsa2004},  \cite{Bayraktar2009}, \cite{Fang2008}, and \cite{Lamberton2008}, among others.  For vested ESOs, the  job termination arrival forces the employee to  exercise immediately. This leads to an optimal stopping problem with forced exit. In terms of the  variational inequality for the option value, this restriction  gives rise to an inhomogeneous term that depends on both the  option payoff and job termination intensity.  When the stock price follows a geometric Brownian motion, we provide the closed-form formulas   for both the  vested and unvested perpetual ESO costs. The optimal exercise threshold can be determined uniquely from a polynomial equation, and it admits an explicit expression in the  case without job termination risk.

 In order to compute  the ESO value and exercise boundary, we apply the  Fourier Stepping Timing  (FST) approach, whereby  the associated  inhomogeneous  PIDVI is simplified by Fourier transform and the optimal exercise  price  is determined  in each time step.  \cite{Jackson2008} apply this approach to price European, American, and barrier options under a number of L\'evy models. For our ESO valuation problem,  the structure of the inhomogeneous PIDVI varies under different job termination intensity specifications. In particular, if the intensity is affine in the (log) stock price, then the  inhomogeneous PIDE for the ESO in the continuation region can be simplified to an inhomogeneous PDE. In the constant intensity case, we  further reduce  the associated PIDE into an ODE. These observations lead to  several  efficient numerical algorithms for valuation.  In all these cases, we compare with   the numerical  results from an implicit-explicit finite difference method for valuing the ESO.  To this end, we refer to the  finite difference  methods for pricing  American options under L\'evy or jump diffusion models, including    \cite{Cont2003,Forsyth03,MadanHirsa2004}, and \cite{Forsyth07}.

 Firms typically expense the granted ESOs according to a fixed schedule, such as  quarterly or annually, but the cost of an ESO changes over time depending on the  stock price movement.   From the firm's perspective, a rise in the  ESO value implies a higher expected cost of compensation as compared to the initially reported value. On the other hand, existing ESOs can be  either exercised voluntarily by the employee, or terminated due to employment termination.   This motivates us to study  (i) the probability that the ESO cost will exceed a given level in the future, and (ii) the contract termination probability.  The ESO cost exceedance  probability bears  similarity to  the loss probability used in classical value-at-risk calculation. The contract termination probability sheds light on the likelihood that the firm will have to pay the employee over a future horizon, from a week to a few years.  We apply Fourier transform based  methods to compute  these probabilities under constant job termination intensity, and we show the connection between our approach and that developed by   \cite{MadanCarr1999} in the computation of  these probabilities. 

The rest of the paper is structured  as follows. In Section \ref{sect-model}, we formulate the ESO valuation framework under \lv price dynamics.   In Section \ref{sec-4}, we discuss the Fourier transform based  numerical  methods for  ESO valuation.  In Section \ref{sect-affine}, we discuss the  valuation problem under stochastic job termination intensity. In Section \ref{sec-5}, we analyze and evaluate the probability of  ESO cost exceedance  and the probability of  contract termination.  In Section \ref{sec-perp}, we  provide closed-form formulas for the vested and unvested perpetual ESO costs under the lognormal stock price model.   Section \ref{sec-6} concludes the paper.



\section{Model Formulation} \label{sect-model}
In the background, we fix  a probability space $(\Omega,\mathcal G,\P)$ satisfying the usual conditions of right continuity and completeness, where $\P$ is the historical probability measure. Let   $(X_t)_{t \ge 0}$ be a L\'{e}vy process, which  admits  the well-known  L\'{e}vy-It\^o decomposition  \citep[ p.119]{Sato1999}:
\begin{align}
X_t=\mu t+\sigma B_t+X^l_t+\lim_{\epsilon \searrow 0}  X^{\epsilon}_t, \quad X_0= 0,
\end{align}
where $B$ is a standard Brownian motion  under $\P$, and the jump terms are given by
\begin{align}
X^l_t&=\int_{|y| \ge 1,s \in [0,t]}yJ(dy,ds),\label{JJ1}\\
  X^{\epsilon}_t&=\int_{\epsilon \le |y| < 1,s \in [0,t]}y(J(dy,ds)-\nu(dy)\,ds)=\int_{\epsilon \le |y| < 1,s \in [0,t]}y\tilde J(dy,ds).\label{JJ2}
\end{align}
The characteristic triplet  ($\mu, \sigma ^2, \nu$)  of  $X$ consists of the  constant  drift  $\mu$ and volatility   $\sigma$, along with the L\'{e}vy measure  $\nu$. In \eqref{JJ1} and \eqref{JJ2},   the Poisson random measure $J(dy, ds)$ counts the number of
jumps of size $y$ occurring at time $s$, and $\tilde J$ is the associated
compensator.

The characteristic exponent $\Psi(\omega)$ of  $X$ is given
by the L\a'{e}vy-Khintchine formula  \citep[ p.119]{Sato1999}:
\begin{equation} \Psi(\omega)=i\mu \omega-{1 \over 2}\sigma ^2
{\omega}^2+\int_{\mathbb {R}}(e^{i\omega y}-1-iy \omega
\indic{|y|<1})\,\nu(dy), \quad \omega \in \mathbb{C}.\label{Psi}
\end{equation}
With this, the characteristic function of $X_t$ is
$\phi_{X_t}(\omega)=e^{\Psi(\omega)t}$. We denote
$\mathbb{F}^X=(\mathcal{F}^X_t)_{t \ge 0}$ as the filtration generated
by $X$.

 In Table \ref{lv}, we summarize the \lv densities and characteristic exponents for several well-known \lv models that will be used in this paper.  In the  GBM model \citep{BS73}, the L\'evy density is absent as the stock price has no jumps.  The \cite{Merton_jump1976} jump diffusion model features normally distributed  jump sizes, while  the \cite{Kou2002}  model assumes  a double exponential distribution for the jump sizes. In contrast, under the  Variance Gamma (VG) model (\cite{MadanCarr1998}) and  CGMY model (\cite{MadanCarr2002}), the stock price follows a pure jump process with infinite activity.

\begin{table}[th]\begin{center}\begin{tabular}[]{lll}
\hline
  {Model}  &  {L\'{e}vy Density $\nu(dy)$} & {Characteristic Exponent $\Psi (\omega)$}  \\
\hline\hline
GBM & N/A & $i \mu \omega-{{\sigma ^2 \omega^2} \over 2}$  \\
Merton\!  &${\alpha \over \sqrt {2\pi \tilde {\sigma} ^2}}e ^{-{1 \over 2}((y-\tilde \mu)/ \tilde \sigma)^2}$  & $i \mu \omega-{{\sigma ^2 \omega^2} \over 2}+ \alpha(e ^{i \tilde \mu \omega-{{\tilde \sigma ^2 \omega^2} \over 2}}-1)$ \\
Kou & $\alpha \big ({p \eta _+ }e ^{{-y  \eta _+}} \indic{y>0}\!\!+{(1-p)  {\eta _{-}} }e ^{{-|y| \eta_{-}}} \indic{y<0})$ & $i \mu \omega-{{\sigma ^2 \omega^2} \over 2}+\alpha({p \over {1-i\omega / \eta_+}}+{{1-p} \over {1-i\omega / \eta_ -}}-1 \big )$\\
VG & ${1 \over {\kappa |y|}}e^{b_1y - b_2 |y|}$ & $- {1 \over \kappa} \log(1-i {\tilde \mu} \kappa \omega + {{{\tilde \sigma} ^2 \kappa \omega ^2} \over 2})$ \\
CGMY\!  & ${C \over |y|^ {1+Y}}\big (e^{-G|y|}\indic {y<0}+e^{-My}\indic{y>0} \big )$ & $C \Gamma (-Y)\big [ (M-i\omega)^ Y\!\!\! -M ^ Y\!\! \!+\!(G+i\omega)^ Y \!\!-\!G ^ Y \big ]$\\
\hline
\end{tabular}\end{center}
 \caption{  \small {L\'evy densities and characteristic exponents for some well-known L\' {e}vy processes}.
 In all these models, $\mu$ and $\sigma$ are the drift and volatility of the Brownian Motion under $\mathbb{P}$. For the Merton   and Kou models, $\alpha $ denotes the jump intensity. In the Merton model, $\tilde \mu$ and $\tilde{\sigma}^2$ are the mean and variance of the IID normally distributed jumps. In the Kou model, $p$ (resp. $(1-p)$) is the probability of positive (resp. negative) exponential jumps. In the VG model, $b_1 = {{\tilde \mu} / {\tilde \sigma ^2}}$, $b_2={\sqrt{\tilde \mu ^2+2\tilde \sigma ^2/\kappa} / {\tilde \sigma ^2}}$, and in the CGMY model,  ${C}, G, {M} >0,  Y \le 2$.  } \label{lv}
\end{table}

The company stock price is  modeled by  an exponential L\'{e}vy process \[S_t=S_0e^{X_t}, \quad t\ge 0,\] with constant initial stock price $S_0 >0$.
In addition, we assume positive constant interest rate $r$ and non-negative dividend rate $q$. For ESO valuation, we  work with a risk-neutral pricing measure $\Q$ such that
\begin{equation}
\mathbb E^{\Q} \{\,e^{X_1}\}=e^{r-q}\quad \Leftrightarrow \quad \hat\Psi(-i)=r-q,\label{RNcond}
\end{equation}
where $\hat{\Psi}$ is given in \eqref{Psi} with  $\mu$  replaced by
\begin{align}
\hat{\mu}=r-q-{{\sigma ^2} \over 2}- \int_{\mathbb {R}}(e^{y}-1-y\indic{|y|<1}) \,\hat\nu(dy),\label{muvalue}
\end{align}and $\hat{\nu}$ is the L\'evy measure under $\mathbb{Q}$.

\subsection{Payoff Structure}
 Figure \ref{Fig_ESOpayoff} illustrates the  payoff structure of an ESO with strike $K$,  vesting period of $t_v$ years, and expiration date $T$.     During the vesting period, the ESO is not exercisable and is forfeited if the employee leaves the firm. As soon as the option is \emph{vested} (at or after time $t_v$), the employee can exercise the option at any time prior to the expiration date $T$, but will  also be forced to exercise immediately upon job termination. We model the employee's job termination time $\tau^\lambda$ by an exponential random variable with rate parameter $\lambda\ge 0$, and assume that $\tau^\lambda$  and $X$ are independent.  The case of stochastic job termination will be discussed in Section \ref{sect-affine}.

\begin{figure}[thb]
\begin{center}
\includegraphics[width=4.5in, angle=0]{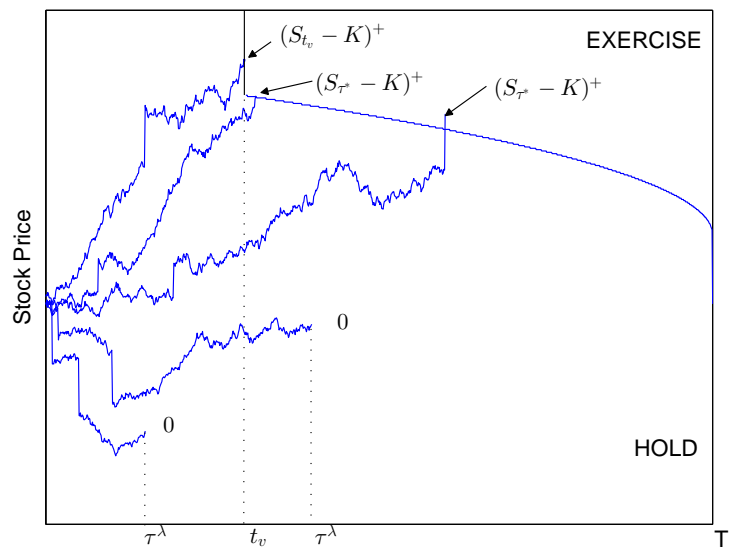}
\caption{\small{{{ESO payoff structure}. From bottom to top:  (\romannumeral1)  The employee leaves the firm during
the vesting period, resulting in forfeiture of the ESO. (\romannumeral2) The employee is forced to exercise the  vested ESO early due to job termination.  (\romannumeral3) The stock price path jumps across the exercise boundary  after vesting, so the employee exercises the ESO immediately. (\romannumeral4) The stock price reaches the  exercise boundary (without jump) after vesting, and the option is exercised there. (\romannumeral5) The employee exercises the ESO at the end of vesting. Along each stock price path, the  vertical line segments depict jumps in the stock price. }}}\label{Fig_ESOpayoff}
\end{center}
\end{figure}

\newpage

\subsection{ESO Cost}
The value of a vested ESO at time $t \in [t_v, T]$  is given by
\begin{align}
\Cvest(t,x) &=\sup_{ \tau \in \setT_{t,T}} \E^{\Q}_{t,x}  \biggl\{\, e^{-r(\tau \wedge
\tl-t)}( S_0e^{{X_{\tau \wedge \tl}}} -K)^+
\,\biggr\}\label{Cost_vest}\\
&= \sup_{ \tau \in \setT_{t,T}}\E^{\Q}_{t,x}  \biggl\{\, e^{- (r+\lambda)(\tau-t)} ( S_0e^{X_{\tau }}
-K)^+ + \int_{t}^{\tau} e^{-(r+\lambda) (u-t)} \lambda ( S_0e^{X_u} -K)^+
du \,\biggr\},\label{Cost_vest2}
\end{align}
where  $\setT_{t,T}$ is the set of $\mathbb F^{X}$ stopping times taking values in $[t,T]$, and $\E^{\Q}_{t,x}\{\cdot\}$ denotes the conditional expectation with $X_t=x$. In other words, after the vesting period, the employee faces an  optimal stopping problem similar to that for an American call option, but is subject to  forced early exercise due to sudden job termination. From (\ref{Cost_vest2}), we can also interpret the vested ESO as an American call option with a cash flow stream of $\lambda (S_0 e^{X_t} -K)^+$   up to the exercise time $\tau ^{\lambda}$. Using the ESO payoff structure, it is straightforward to show that  the  vested ESO cost  $C(t,x)$ is increasing and convex in $x$ for every $t \in [t_v,T)$, and is decreasing in $t$ for every $x \in \mathbb R$.

During the vesting period, the ESO is forfeited if the employee leaves the firm. Hence, given the ESO is still alive  at time $t\le t_v$,  the value of an \emph{unvested} ESO is
\begin{align}
\Cunvest(t,x) &= \E^{\Q}_{t,x}  \biggl\{ \, e^{-r(t_v-t)}
\Cvest(t_v,X_{t_v}) \indic{\tau^{\tilde \lambda} > t_v}
\,\biggr\}\\
&= \E^{\Q}_{t,x}  \biggl\{ \, e^{-(r+\lambda)(t_v-t)}
\Cvest(t_v,X_{t_v}) \,\biggr\}. \label{Cost_unvest}
\end{align} The vesting provision prohibits the employee from exercising the option even if the ESO happens to be in the money during $[0, t_v)$.

The valuation of a  vested  ESO  leads to the analytical and numerical studies of an  inhomogeneous partial integro-differential  variational inequality (PIDVI). To this end, we first define the infinitesimal generator of  $X$ under $\Q$
\begin{equation}\label{Fr}
\hat{\mathcal{L}}f(x)=\hat{\mu} f'(x)+{{\sigma ^2} \over {2}}f''(x)+\int_{\mathbb {R}\backslash{\{0\}}} \big(f(x+y)-f(x)-yf'(x)\indic{|y|<1}\big)\hat\nu(dy),
\end{equation}
with $\hat{\mu}$ given in \eqref{muvalue}.  For the  vested ESO cost,     job termination risk gives rise to an inhomogeneous term in the PIDVI, namely,
\begin{align}
\min\biggl \{-(\partial _t+\hat {\mathcal{L}})C+(r+\lambda)C-\lambda (S_0e^x-K)^+,~ C(t,x)-(S_0e^x-K)^+\biggl \}=0,
\label{VI}
\end{align}
for $(t,x)\in (t_v, T)\times \R$, with  terminal condition $C(T,x)=(S_0e^x-K)^+$, for $x \in \mathbb R$.

For the unvested ESO cost, we set the terminal condition at time $t_v$ by matching it with the vested ESO cost, namely,  $\tilde C(t_v,x)=C(t_v,x)$,  for $x \in \mathbb R$. During the vesting period,   the unvested ESO cost satisfies the  partial integro-differential equation (PIDE)
\begin{equation}
(\partial _t+\hat {\mathcal{L}})\tilde C-(r+\tilde \lambda)\tilde C=0, \qquad (t,x)\in [0,t_v)\times \R.\label{PIDECtil}
\end{equation}

When the vesting period coincides with maturity,   the ESO becomes European-style  as no early exercise is permitted. Setting $t_v=T$ yields the  \emph{European ESO} cost
\[\tilde C^E(t,x) =\E^{\Q}_{t,x}  \biggl\{ \, e^{-r(T-t)} (S_0e^{X_T} - K)^+ \indic{\tau^{\tilde \lambda} > T}
\,\biggr\}, \quad 0 \le t\le T. \]
This cost function also satisfies the PIDE \eqref{PIDECtil}.

%

\subsection{Exercise Boundary}
 For a vested ESO, the holder's exercise strategy can be described by  the optimal exercise boundary $t \mapsto s^*(t)$  that divides the  domain $[t_v,T) \times \mathbb R$   into the continuation region $\mathcal {C}$ and   exercise region $\mathcal {S}$, defined by
 \begin{align}
\mathcal {C}&=\{(t,s) \in [t_v,T) \times \mathbb R_+, s <s^*(t)\},\label{regionC}\\
\mathcal {S}&=\{(t,s) \in [t_v,T) \times \mathbb R_+, s \ge s^*(t)\},\label{regionS}
\end{align}  where \begin{equation}
s^*(t):=\sup\{s \ge 0\, |\, C(t,x)>(s-K)^+\}, \quad \text{ for }\,  t \in [t_v,T).\label{sstar}
\end{equation}

If $s^*(t)<+\infty$, then we have $C(t,x)>(s-K)^+$   for $s<s^*(t)$, and $C(t,x)=(s-K)^+$ for $s \ge s^*(t)$, for $s =S_0e^x \in \big(s^*(t),+\infty \big)$, due to the convexity and positivity of $C(t,x)$.  Since,  for  every fixed $x\in \R$, $C(t,x)$ is decreasing in $t$,  the optimal exercise boundary   $s^*(t)$ must be decreasing in $t$ in view of \eqref{sstar}.  As for the impact of  job termination risk, we have the following result:

\begin{proposition}\label{proplast}
  A higher post-vesting job termination intensity decreases the costs of  vested and unvested ESOs, and lowers the optimal exercise boundary.
\end{proposition}

We provide a proof in the Appendix \ref{Appendix1}. A similar result has been established under the GBM model by   \cite{LeungSircarESO_MF09}.  We remark that the cost reduction effect of the post-vesting job termination intensity holds   with and without dividends. However, with  $q=0$, the job termination does not affect the employee's voluntary exercise timing since it is optimal not to  exercise early voluntarily regardless of  job termination risk.  In this case, the value of an  American  ESO equals to  that  of a European  ESO, and  the variational inequality (\ref{VI}) is  simplified to a PIDE.    On the other hand, a higher pre-vesting job termination intensity can reduce the unvested ESO cost, but has no impact on the vested ESO value or the post-vesting exercise strategy.

\begin{remark}  In our paper,   the optimal exercise boundary is computed based on  the risk-neutral pricing measure. In practice, the ESO holder is likely    unable to perfectly hedge the ESO exposure. In addition, the ESO holder may opt  to exercise early  due to other exogenous factors, such as liquidity risk or  need for diversification.  To this end, one can develop  a  reduced form or  intensity-based approach  by treating   the ESO exercise timing as fully exogeneous, and  calibrating to observed exercise behaviors.  
\end{remark}

\section{Fourier Transform Method for ESO Valuation}\label{sec-4}
For ESO valuation, we now discuss a numerical approach for solving the PIDVI \eqref{VI} based on Fourier transform.  We first  state the definition and some basic properties of Fourier transform. For any  function $f(x)$, the associated  Fourier transform  is defined by
\begin{equation}
\mathcal {F}[f](\omega)= \int _{-\infty} ^{\infty} f(x)e^{-i\omega x}dx,
\end{equation} with angular frequency  $\omega$  in radians per second.
In turn, if we denote by $\hat f(\omega)$ the Fourier transform of $f(x)$, then its inverse Fourier transform is
\begin{equation}
\mathcal {F} ^{-1}[\hat f](x)= {1 \over 2\pi}\int _{-\infty} ^{\infty} \hat f(\omega)e^{i\omega x}d\omega.
\end{equation}
As is well known,  the Fourier transform of derivatives satisfies \begin{equation}\label{FTprop1}
\mathcal {F}[\partial _x ^nf](\omega)=i\omega\mathcal {F}[\partial _x ^{n-1}f](\omega)=\ldots=(i\omega)^n\mathcal {F}[f](\omega).
\end{equation}
Applying \eqref{FTprop1} to \eqref{Fr}, we have
\begin{align}
\mathcal{F}[\hat{\mathcal L} f](\omega)
&=\{i \hat{\mu} \omega-{1 \over 2} \sigma ^2 \omega^2+ \int _\R (e ^{i\omega \hat \mu}-1-i y \omega 1_{\{|y|<1\}})\hat \nu(dy)\}\mathcal F[f](\omega)  \,\notag\\
&=\hat\Psi(\omega)\mathcal{F}[f](\omega),\label{FFFtt}
\end{align}
where $\hat\Psi$ is the characteristic exponent under $\mathbb{Q}$.

In the continuation region,  the vested ESO cost $C(t,x)$ satisfies   the inhomogeneous PIDE
\begin{align}
-(\partial _t+\hat {\mathcal{L}})C+(r+\lambda)C-\varphi(x)=0,
\label{PIDE of vested ESO}
\end{align}
where $\varphi(x):=\lambda(S_0e^x-K)^+$.   


 An application of  Fourier transform  to   \eqref{PIDE of vested ESO} yields
\begin{equation}
\partial_t\mathcal{F}[C](t,\omega)+\big (\hat\Psi(\omega)-r-\lambda \big )\mathcal{F}[C](t,\omega)=-\mathcal{F}[\varphi](\omega) .
\label{ODE of vested ESO}
\end{equation}
Therefore, the original   inhomogeneous PIDE is  transformed into an   inhomogeneous  ODE (\ref{ODE of vested ESO})  satisfied by $\mathcal{F}[C](t,\omega)$, a function of time $t$ parametrized by $\omega$. Given the value of $\mathcal F[C]$ at any time $t_2\le T$, we have at an earlier time $t_1$  that
\begin{equation}
\mathcal{F}[C](t_1,\omega)=\mathcal{F}[C](t_2,\omega)e^{(\hat\Psi(\omega)-r-\lambda)(t_2-t_1)}+({\mathcal{F}[\varphi](\omega) \over \hat\Psi(\omega)-r-\lambda})(e^{(\hat\Psi(\omega)-r-\lambda)(t_2-t_1)}-1), \quad t_1<t_2.
\end{equation}
By inverse Fourier transform, we recover the vested ESO cost in the continuation region
\begin{equation}
C(t_1,x)=\mathcal{F}^{-1}[ \mathcal{F}[C](t_2,\omega)e^{(\hat\Psi(\omega)-r-\lambda)(t_2-t_1)}] (x)+\mathcal{F}^{-1}[({\mathcal{F}[\varphi](\omega) \over \hat\Psi(\omega)-r-\lambda})(e^{(\hat\Psi(\omega)-r-\lambda)(t_2-t_1)}-1)](x).
\end{equation}

Since the ESO is early exercisable, we  need to compare the vested ESO value with the payoff from immediate exercise. Precisely,  we partition the time interval $[t_v, T]$ into  $ t_m, m=M, M-1, \ldots, 1$, then we iterate backward in time with
\begin{align}
\begin{cases}
{ C(t_{m-1},x)
=\mathcal{F}^{-1}[\mathcal{F}[C](t_m,\omega)e^{(\hat\Psi(\omega)-r-\lambda)(t_m -t_{m-1})}
+{\mathcal{F}[\varphi](\omega) \over \hat\Psi(\omega)-r-\lambda}(e^{(\hat\Psi(\omega)-r-\lambda)(t_m -t_{m-1})}-1)](x),}\\
{C(t_{m-1},x) = \max \{  C(t_{m-1},x), (S_0e^x- K)^+  \},}
\end{cases}
\end{align}
where $t_M=T$ and $t_0=t_v$.

For numerical  implementation, we  discretize  the original domain  $\Omega =[t_v,T] \times \R$ into  a finite grid: $\{(t_m,x_n):m=0,1,\ldots,M, n=0,1,\ldots,N-1 \}$, where $t_m=t_v+m \Delta t$, and $x_n=x_{min}+n \Delta x$, with $\Delta t=(T-t_v)/M$ and $\Delta x=(x_{max}-x_{min})/(N-1)$.  As most ESOs are  granted at the money, it is natural to set the upper/lower price bounds $x_{min}=-x_{max}$ equidistant from zero. With   $x_{max}$,  $M$, and $N$  fixed, we apply the Nyquist critical frequency $\omega_{max}={\pi / \Delta x}$ and set $\Delta \omega=2\omega_{max}/N$.

The continuous Fourier transform is approximated by the discrete Fourier transform (DFT)
\begin{align}
\mathcal F[C](t_m, \omega _n) &\approx \sum _{k=0} ^{N-1} C(t_m,x_k)e^{-i \omega _n x_k} \Delta x =\alpha _n  \sum _{k=0} ^{N-1} C(t_m,x_k)e^{{{-i nk} \over N}},\label{FFT1}
\end{align}
with $\alpha _n = e^{-i \omega _n x_{min}} \Delta x$.  In \eqref{FFT1}, we evaluate  the sum  $\sum _{k=0} ^{N-1} C(t_m,x_k)e^{{{-i nk} \over N}}$   using the Fast Fourier Transform (FFT) algorithm. The corresponding Fourier inversion is conducted by inverse FFT, yielding the vested ESO cost $C(t_m, x_n)$. Note that  the coefficient $\alpha _n$ will be cancelled in the process.

After computing the vested ESO values,    the unvested ESO cost is given by
\begin{align}\label{unvested}\tilde C(t,x)&=\mathcal{F}^{-1}\big [\mathcal{F}[\tilde C](t_v,\omega)e^{(\hat\Psi(\omega)-r-\tilde \lambda)(t_v-t)}\big ](x),\quad \text{ for } ~t\le t_v,
\end{align} where  $\tilde C(t_v,x)= C(t_v,x)$. Again, the Fourier and its inversion are implemented via FFT.
%

We now  provide some numerical results to illustrate  the application of the pricing methods discussed above.   In Table \ref{Tab_constant}, we summarize the ESO costs for  different vesting periods $t_v = 0, 2, 4$. As a  numerical  check, we compare with the ESO costs computed  from a finite difference  method (see Appendix \ref{app_fdm}), and observe that  the two numerical methods return very close values.

\begin{table}[h!]
\centering
\begin{tabular}{c|cc|cc|cc}
\hline
Model & \multicolumn{2}{c}{ $t_v = 0$} & \multicolumn{2}{c}{ $t_v=2$} &\multicolumn{2}{c}{ $t_v=4$}
\\ \cline{2-7}
 & FST  & FDM & FST  & FDM & FST & FDM \\ \hline\hline
GBM & 1.3736 &1.3730 & 1.3822  &1.3816 & 1.2365  &1.2360\\
Merton  & 1.4820  & 1.4803& 1.4899  &  1.4887& 1.3313 &1.3306 \\
Kou  & 1.4566  & 1.4558 & 1.4648  &  1.4646&1.3091  &1.3104\\
VG & 1.5584  &1.5595 &1.5816  & 1.5811&  1.4131 &1.4139  \\
CGMY & 1.8409    & 1.8411 & 1.8532 &1.8535& 1.6484  & 1.6490 \\ \hline
\end{tabular}  \caption{\small{ESO cost comparison. For each vesting period $(t_v = 0, 2, 4)$, the three columns  contain ESO costs computed, respectively, by FSTC,  FSTG, and  the finite difference method. Common parameters: $S_0=K=10, r=0.05, q=0.04, \lambda =0.2, \tilde \lambda =0.1, T=8$. In the GBM model, $\sigma=0.2$. In the Merton  model, $\sigma=0.2,\alpha = 3, \tilde \mu=0.02, \tilde \sigma=0.045$. In the Kou  model, $p=0.5, \sigma=0.2,\alpha =3,\eta_+=50, \eta_-=25$. In the Variance Gamma model, $\tilde \mu=-0.22, \tilde \sigma=0.2, \kappa=0.5$.  In the CGMY model, $C=1.1, G=10, M=10, Y=0.6$. The grid sizes for FST methods are the same, with  $x_{max} =6$, $M=2048$, $N=32768.$ }} \label{Tab_constant}
\end{table}

%
%

Figure \ref{fig_ejob} illustrates how the optimal  exercise boundary changes with respect to  job termination intensity $\lambda$ and stock price jump intensity $\alpha$.  As $\lambda$ increases from $0.1$ to $0.3$,  the optimal exercise boundary is lowered. As the stock price jump intensity increases,  the optimal exercise boundary moves upward.  Also, we remark that in all cases the exercise boundary is decreasing in  time $t$.

The cost impact of job termination intensity and vesting period  is demonstrated in Figure \ref{fig_cjob}. As suggested by Proposition \ref{proplast}, a  higher post-vesting job termination intensity reduces the ESO cost. Also, when   the post-vesting and pre-vesting  job termination rates are the same, the ESO cost decreases  as vesting period lengthens (see  Figure \ref{fig_cjob} (left)). However, if  the post-vesting job termination rate ($\lambda = 0.2$) is higher than the pre-vesting rate ($\tilde \lambda = 0.1$), then it is possible that the ESO can first  increase with vesting period.


\begin{figure}[h]
  \includegraphics[width=3.5in,height=2in]{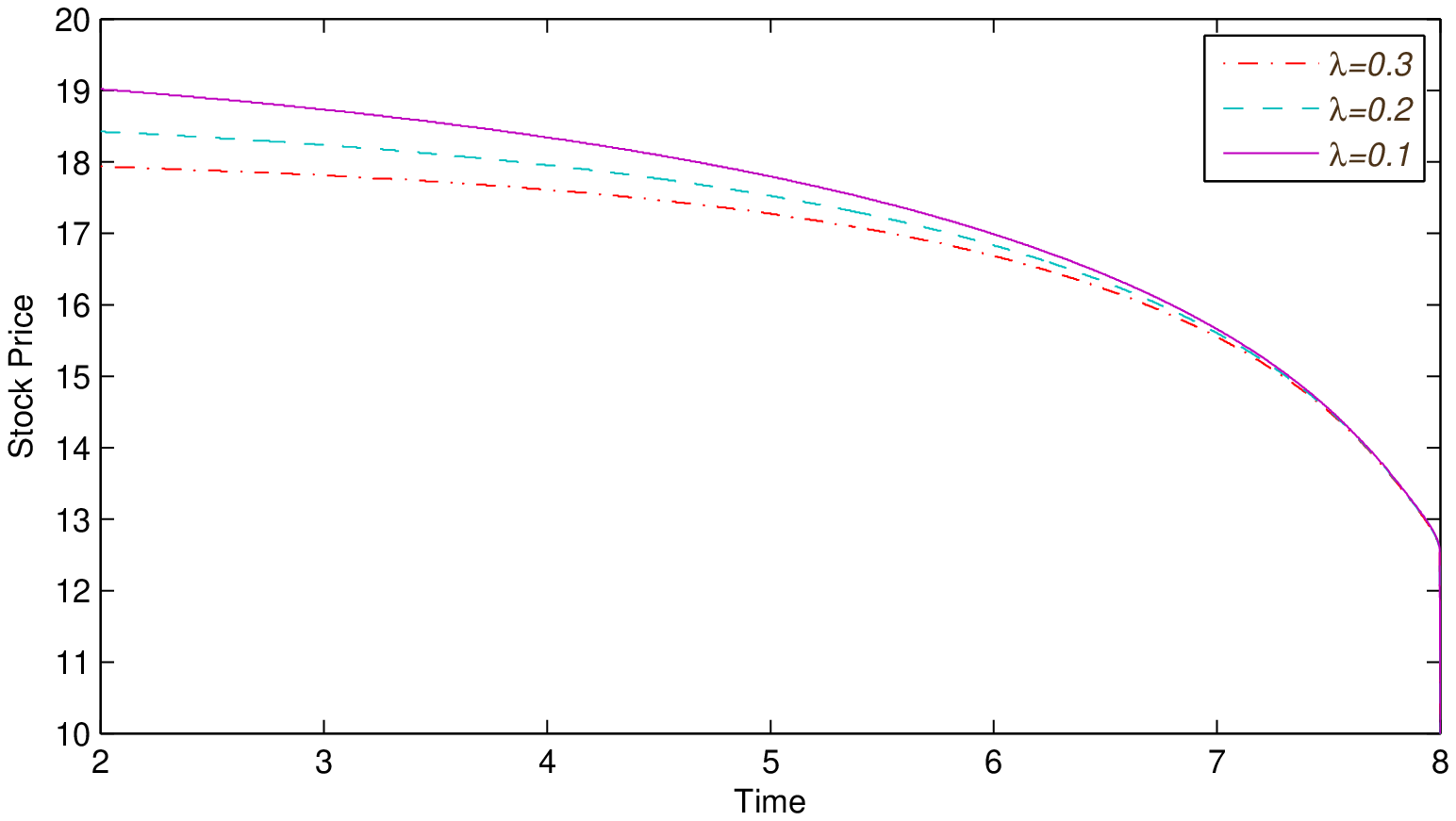}
    \includegraphics[width=3.5in,height=2in]{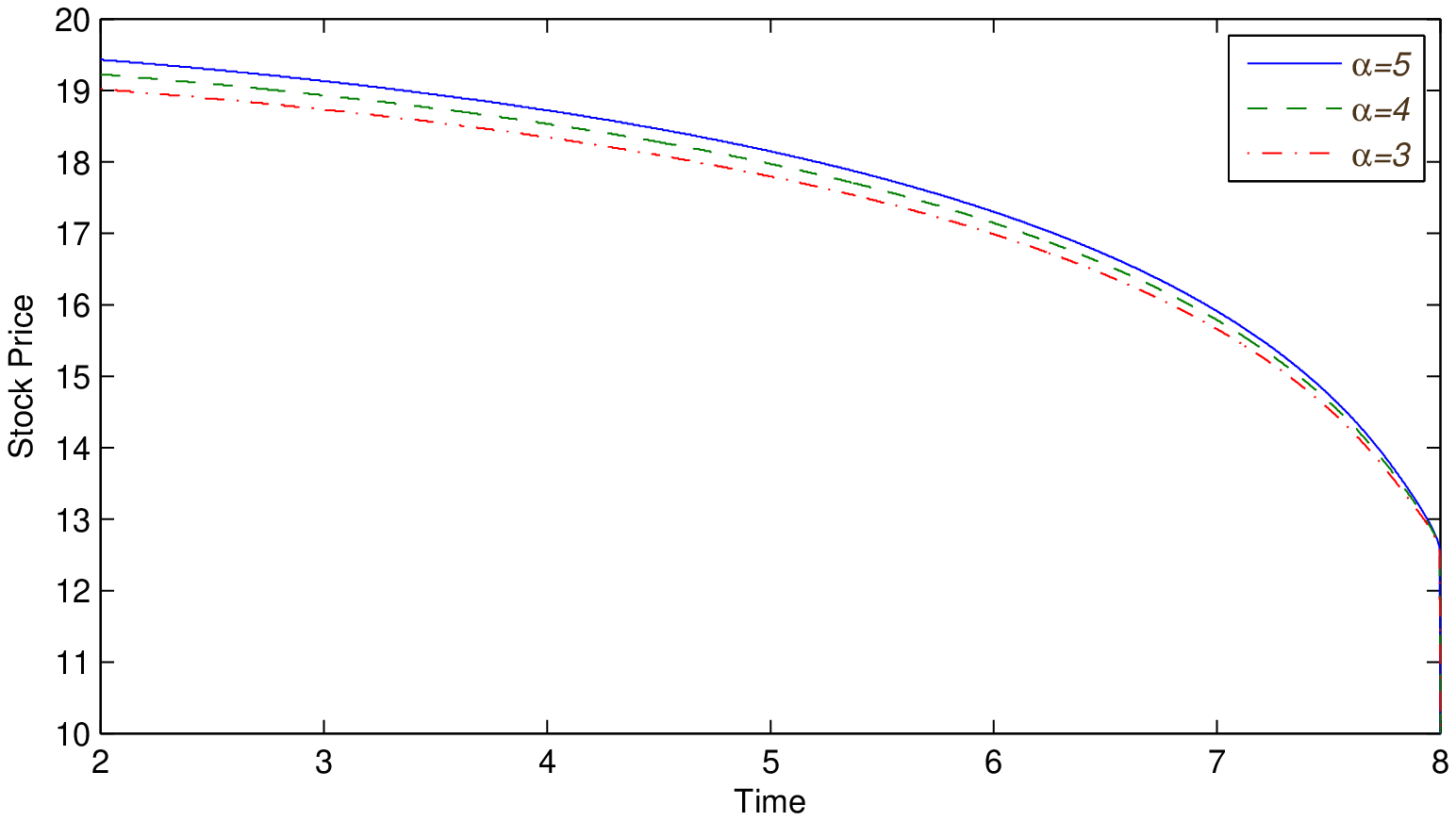}
   \caption{\small {(Left)   ESO exercise boundaries for different job termination rates.   (Right)  ESO exercise boundaries for different  stock price jump intensities. This example is based on the Kou model with  parameters: $S_0=K=10, r=0.05, p=0.5, \sigma=0.2, q=0.04, \eta_+=50, \eta _-=25,  \tilde \lambda = 0.1, T=8$.} }\label{fig_ejob}
\end{figure}

\begin{figure}[h]
    \includegraphics[width=3.5in,height=2in]{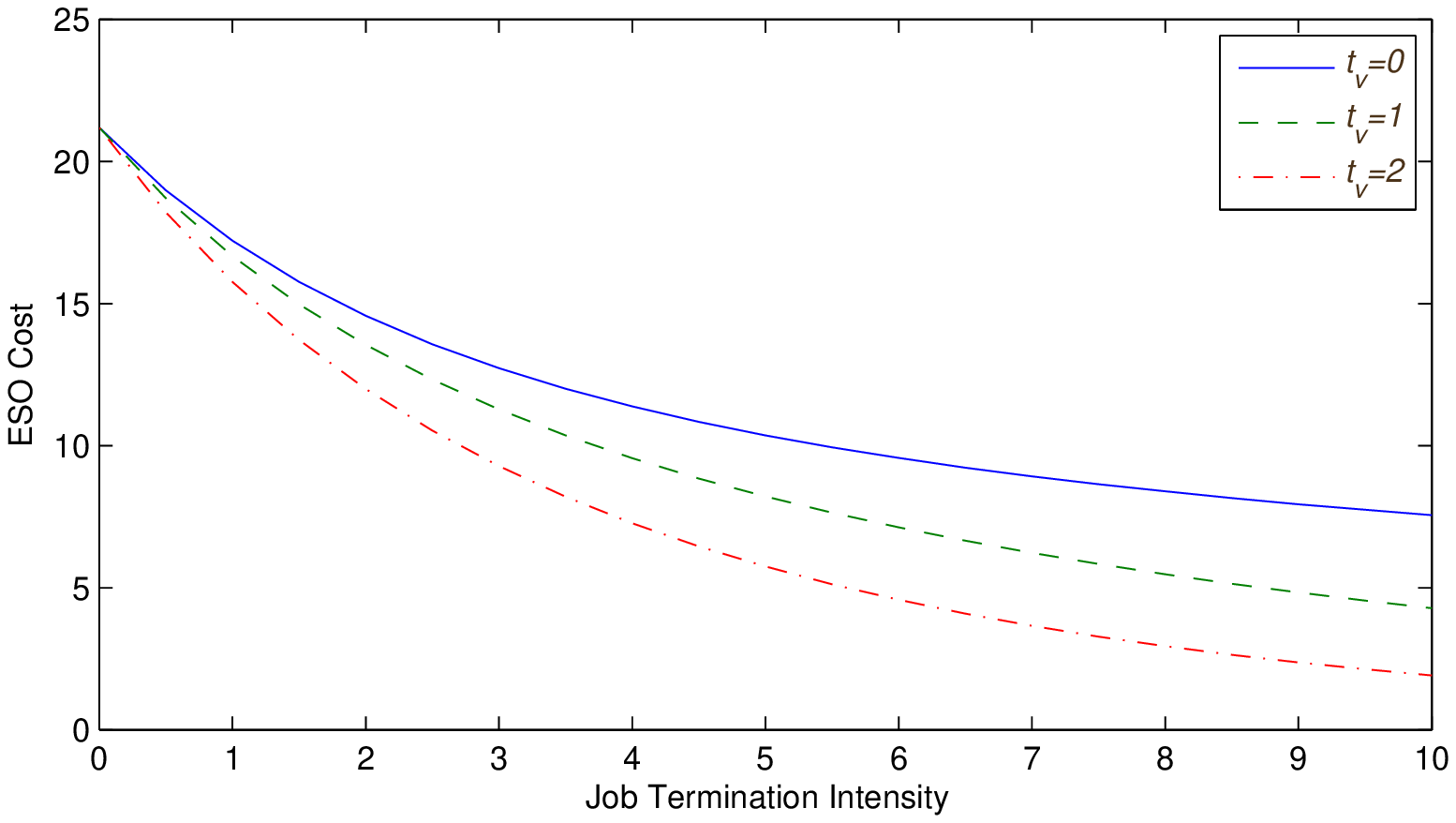}
         \includegraphics[width=3.5in,height=2in]{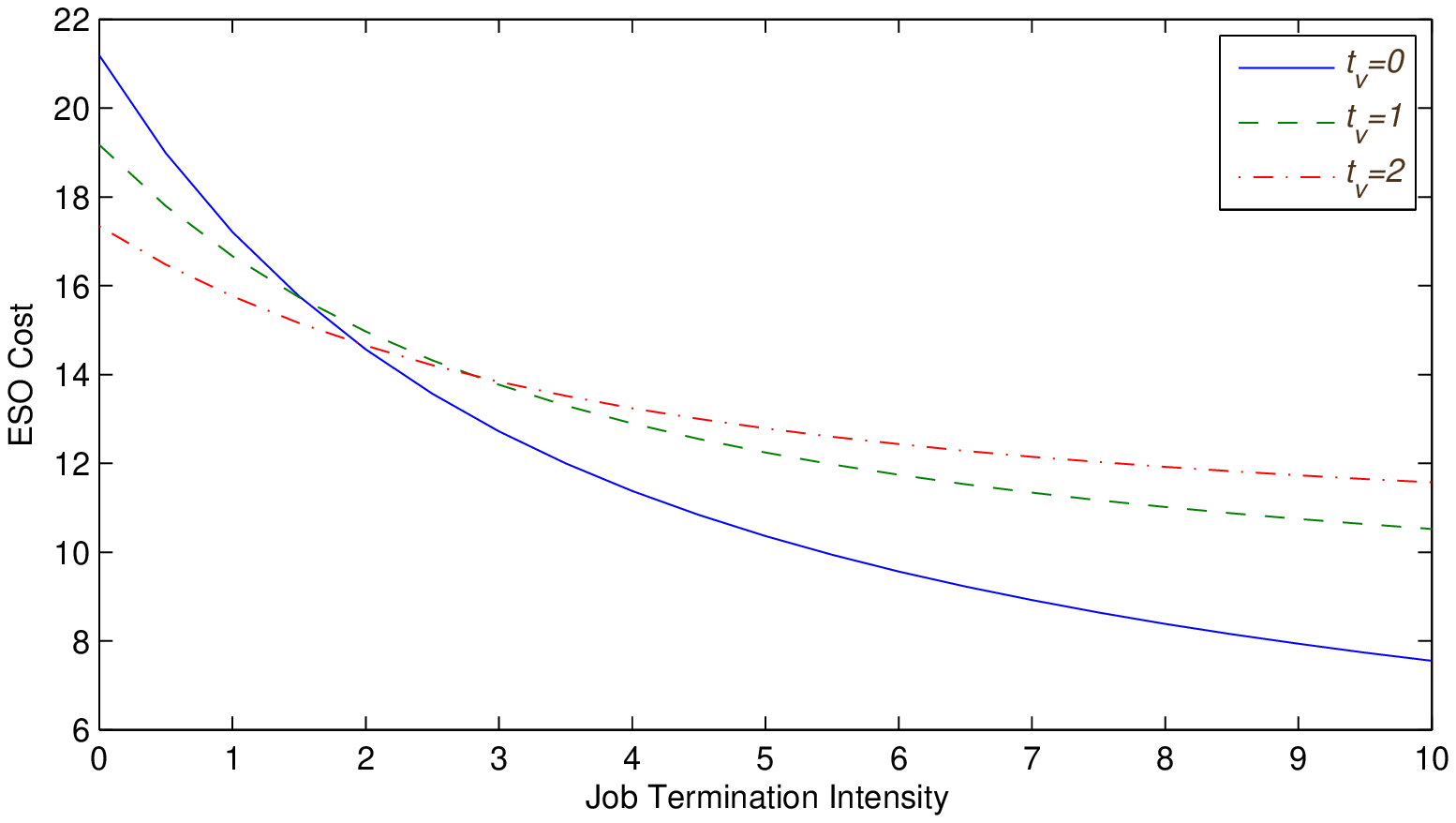}
 \caption{\small{(Left) The ESO cost decreases  as vesting period $t_v$ lengthens or  as post-vesting job termination rate $\lambda$ increases.  (Right) With  $\lambda = 0.2$  and $\tilde \lambda = 0.1$, longer  vesting  can increase the ESO cost.  This example is based on the Kou  model with  parameters: $S_0=K=10, r=0.05, p=0.5, \sigma=0.2, q=0.04, \eta_+=50, \eta_-=25, T=8$.}}\label{fig_cjob}
 \end{figure}


\begin{remark} The model proposed by  \cite{Cvitanic2008} assumes that the ESO holder will voluntarily exercise as soon as  the log-normal stock price  reaches an upper exogenous  barrier. In addition, they  also incorporate a  vesting period and  constant job termination intensity.  Our current framework can also be adapted to their model. Precisely, one can numerically solve the PIDE
\begin{align}
(-\partial _t+\hat{\mathcal{L}})C+(r+\lambda)C-\lambda(S_0e^x-K)^+=0,\quad
\end{align}
with  the modified boundary condition $C(t,x) -C(t,L(t))=0$, for $ (t,x)\in (t_v, T)\times (\log(L(t)), +\infty)$, and  terminal condition $C(T,x)=(S_0e^x-K)^+$. Here,  the function $L(\cdot)$ is a given  exponentially decaying barrier. An ESO cost comparison  is provided in  Table \ref{table_Cvit} below.
\end{remark}

\begin{table}[h!]
\centering
\small
\begin{tabular}{c|ccc|ccc}
\hline
 & \multicolumn{3}{c|}{$ q =0 $} & \multicolumn{3}{c}{$ q = 0.04 $}\\

  & Barrier ESO & European ESO  & American ESO &Barrier ESO & European ESO  & American ESO
\\ \hline
\hline
L=125  & 22.7792  & 37.5435 & 37.5435 & 15.4209 & 16.5753 & 18.2484 \\
L=150  & 26.8375  & 37.5435  & 37.5435 & 17.4808 & 16.5753 & 18.2484 \\
L=9999 & 37.5450  & 37.5435  & 37.5435  & 16.5751 & 16.5753 & 18.2484 \\ \hline
\end{tabular}
\small
 \caption{ESO cost comparison.  Under the GBM model, the \emph{Barrier ESO} assumes exercise at an exogenous decaying boundary, $L(t) = L e^{at}$ for $t_v \le t\le T$ with $a=-0.02$, as presented in \citep[Table 2]{Cvitanic2008}. In the zero dividend ($q=0$) case, the \emph{European} and \emph{American} ESO costs coincide, and they dominate \emph{Barrier ESO} cost.  Other common parameters: $S_0=K=100, r=0.05, \sigma =0.2, \lambda =0.04, \tilde \lambda =0.04, t_v=3, T=10$. }\label{table_Cvit}
\end{table}
In Table \ref{table_Cvit}, the barrier ESO corresponds to Case D in \cite{Cvitanic2008}, i.e. the optimal exercise boundary is an exponentially decaying curve: $L(t)= Le^{at}$, where $a < 0$. Accordingly, we can see: (1) The ESO cost in \cite{Cvitanic2008} is always underestimated if the ESO holder is allowed to exercise the ESO after the vesting period, because the exogenous exercise boundary is not generally the real optimal exercise boundary for the ESO holder.  (2) In \cite{Cvitanic2008}, when $L$ becomes larger and larger, indicating that the ESO holder is unlikely to exercise the ESO according to exogenous exercise boundary, the ESO cost gets closer to the European ESO cost in our paper. This case is also discussed in \cite{CarrLin00}. (3) When $q = 0$, the value of European ESO is equal to the value of American ESO, and when $ q > 0$, the value of European ESO is less than the value of American ESO. In general, our  algorithm can also be modified to adapt other appropriate  payoff  at job termination. In the special case with zero payoff at job termination, the ESO can be interpreted as being  forfeited at the time of departure. Alternatively, this can be considered as an American call option with default risk and zero recovery.

\section{Stochastic Job Termination Intensity}\label{sect-affine}
As an extension to our ESO valuation model, one   can  randomize the job termination rate  by defining
\[
  \tau^{\lambda} = \inf \bigl\{ \,t \ge 0 : \int_0^t \lambda(s,X_s) \,ds > E \bigr\}, \]
where $\lambda(t,x)$ is a smooth positive deterministic function and $E \sim \exp(1) \perp \Fil^X$.   This Markovian job termination  intensity allows for dependence between $\tau^\lambda$ and $X$ while preserving  tractability   by not increasing the dimension of the PIDVI.   In related ESO studies,   \cite{CarrLin00} and \cite{Cvitanic2008} also consider this Markovian intensity  approach to model job termination and exogenous exercise, though they  do not incorporate optimal voluntary exercise. In particular,  \cite{CarrLin00} consider the job termination  intensity of the form: $\lambda(t,X_t) = \lambda_f + \lambda_e 1_{\{S_0 e^{X_t} > K\}}$. The second  term is to model the  early  exercise due to the holder's exogenous desire for liquidity,  and it is constant if  the ESO is in-the-money and zero otherwise.


In our paper, we    assume  the job termination intensity functions before and after the vesting period   to be  affine in log-price, denoted respectively, by $\lambda(t,x)=ax+b$ and $\tilde \lambda(t,x)= \tilde a x+ \tilde b$, for some constants $a, b, \tilde{a}$, and $\tilde{b}$.  Intuitively, the ESO holder's employment is more at risk when the company stock price is low, so one can let $a$ and $\tilde{a}$ be negative. For implementation, one can select parameters and control the grid size so  that the intensity   remains positive within the truncated log-price interval $[x_{min},x_{max}]$. The affine intensity assumption is utilized in simplifying the associated inhomogeneous PIDE.

The   PIDE   for  the  vested ESO cost in the continuation region is
\begin{equation}
(\partial _t+\hat {\mathcal{L}})C-rC=(ax+b) C+\psi(x),\label{affine1}
\end{equation}
where $\psi(x)=(ax+b)(S_0e^{x}-K)^+$. Differentiating    (\ref{affine1}) w.r.t. $x$, and applying   Fourier transform and \eqref{FTprop1}, we obtain an inhomogeneous  PDE in $t$ and $\omega$, namely,
\begin{equation}
i\omega\big(\partial_t+\big(\hat\Psi(\omega)-r\big)\mathcal{F}[C](t,\omega)\big)=\mathcal{F}[aC+(ax+b)C_x](t,\omega)-i\omega\mathcal{F}[ \psi](\omega),
\label{PDE11 of vested ESO}
\end{equation}
with   terminal condition $\mathcal{F}[C](T,\omega)=\mathcal{F}[(S_0e^{x}-K)^+]$.

Using the following well-known property of  Fourier transform:
\begin{equation}
\mathcal{F}[xC_x](t,\omega)=-\mathcal{F}[C](t,\omega)-\omega \partial _\omega \mathcal{F}[C](t,\omega),
\end{equation} followed by the substitution
 \begin{equation}\label{subb}\mathcal{F}[C](t,\omega) = e^{{-i \over a} \int_{-\infty} ^{\omega} (\hat\Psi(s)-r-b)ds}  H(t,\omega),\end{equation}
 we simplify  (\ref{PDE11 of vested ESO})  to a first-order PDE
\begin{equation}
\partial _t H(t,\omega)+{a \over i}\partial _\omega H(t,\omega)=-e^{{i \over a} \int_{-\infty} ^{\omega} (\hat\Psi(s)-r-b)ds}\mathcal{F}[ \psi](\omega),
\label{tranportation equation}
\end{equation}with  terminal condition $H(T,\omega)=\mathcal{F}[(S_0e^{x}-K)^+](\omega)e^{{i \over a} \int_{-\infty} ^{\omega} (\hat\Psi(s)-r-b)ds}$. Finally, solving   (\ref{tranportation equation}) and unraveling  the substitution, the Fourier transform of $C$ can be expressed as
\begin{equation}
\mathcal F [C(t _1,x)](\omega)=e^{{-i \over a} \int_{-\infty} ^{\omega} (\hat\Psi(s)-r-b)ds}\big(\mathcal{F}[C(t _2,x)](\omega)e^{{i \over a} \int_{-\infty} ^{\omega+{a \over i}\Delta t} (\hat\Psi(s)-r-b)ds}+\int _0 ^{\Delta t}g(\omega-{a \over i}(s-\Delta t))ds \big ),\label{formula1}
\end{equation}
where \[g(\omega)=e^{{i \over a} \int_{-\infty} ^{\omega} (\hat\Psi(s)-r-b)ds}\mathcal{F}[ \psi](\omega), \quad t_v\le t_1< t_2\le T,\quad \Delta t=t_2-t_1.\]

The Fourier transform  \eqref{formula1} allows us to compute the  values of $C(t,x)$ backward in time, starting from expiration date $T$. The numerical implementation of  \eqref{formula1} requires computing   the integral $\int _0 ^{\Delta t}g(\omega-{a \over i}(s-\Delta t))ds$. Within each small  time step  $[t,t+\Delta t]$, we approximate the integral $\int _0 ^{\Delta t}g(\omega-{a \over i}(s-\Delta t))ds$ by summing over a further divided  time discretization, namely,  $ \sum_{k= 0}^{n'-1}g(\omega+{a k \over in'}\Delta t) \Delta t/n' $. Therefore, the solution for vested ESO cost in the continuation region  is given by
\begin{align*}
&C(t _1,x)\\
&=\mathcal F ^{-1}\big [e^{{i \over a} \int_{\omega} ^{\omega+{a \over i}\Delta t}(\hat\Psi(s)-r-b)ds}\mathcal F[C(t _2,x)](\omega+{a \over i}\Delta t)+\sum_{k= 0}^{n'-1}g(\omega+{a k \over in'}\Delta t) {\Delta t \over n'} e^{{-i \over a} \int_{-\infty} ^{\omega} (\hat\Psi(s)-r-b)ds} \big ](x),
\end{align*}
for $ t_v \le t_1 < t_2 \le T$, $x \in \mathbb R $. Again, within each iteration, we impose the condition $C(t _1,x) = \max\{ C(t _1,x), (S_0 e^x-K)^+ \}$ to get the American option value.

As for  the unvested ESO, we have $\tilde C(t_v,x)=C(t_v,x)$ at time $t_v$, and
\[(\partial _t+\hat {\mathcal{L}})\tilde C-(r+(\tilde a x+ \tilde b))\tilde C=0,\] for $(t,x) \in [0,t_v)  \times \R$.  At time $t< t_v$,  the unvested ESO cost is computed by
\begin{align}
\tilde C(t,x)
=\mathcal F^{-1}\big [\mathcal{F}[\tilde C(t_v,x)](\omega+{\tilde a \over i}( t_v-t ) )e^{{i \over \tilde a} \int_{\omega} ^{\omega+{\tilde a \over i}(t_v -t)} (\hat\Psi(s)-r-\tilde b)ds} \big ](x).
\end{align}


%


\begin{remark}\label{sect-general}
  Suppose the pre-vesting and post-vesting job termination intensities, $\lambda$ and $\tilde{\lambda}$, are   positive bounded  functions of $x$ and $t$. The  Fourier transform of the ESO cost satisfies
\begin{equation}
\partial_t\mathcal{F}[C](t,\omega)+\big(\hat\Psi(\omega)-r\big )\mathcal{F}[C](t,\omega)-\mathcal{F}[\lambda C](t,\omega)=-\mathcal{F}[\xi](\omega),
 \label{ODE123}
\end{equation}
where $\xi(t,x)= \lambda(t,x)(S_0e^x-K)^+$. In order to solve  this  ODE,  one can apply an  \emph{explicit} scheme to the term $\mathcal{F}[\lambda C](t,\omega)$, and apply \emph{implicit} scheme to the other terms. Therefore, given the value of $C(t_2,x)$ at time $t_2>t_1$, we compute the value of $C(t_1,x)$  by
\begin{align}
C(t_1,x)&=\mathcal{F}^{-1}\big [\mathcal{F}[C](t_2,\omega)e^{(\hat\Psi(\omega)-r)(t_2-t_1)}\big]\notag  \,
+\mathcal{F}^{-1}\big[{\mathcal{F}[\xi](\omega)-\mathcal{F}[\lambda C](t_2,\omega)\over \hat\Psi(\omega)-r}(e^{(\hat\Psi(\omega)-r)(t_2-t_1)}-1)\big ].
\end{align}
With this, we can compute the vested ESO cost  by iterating  backward in time and comparing with the payoff from immediate exercise.   Finally, we remark that this implicit-explicit algorithm is also applicable  when the job termination  intensity is  constant or  affine.
\end{remark}

In Table \ref{Tab_costaffine}, we compute the ESO cost with affine   pre-vesting  and post-vesting job termination rates  using the  FST method in Section \ref{sect-affine} and that in Remark  \ref {sect-general}, along with the finite difference method in Appendix \ref{app_fdm}. We    observe that ESO costs for  different vesting periods from  these three different methods are very close.

\begin{table}[h!]
\centering
\begin{tabular}{c|ccc|ccc|ccc}
\hline
Model & \multicolumn{3}{c}{ $t_v = 0$} & \multicolumn{3}{c}{ $t_v=2$} &\multicolumn{3}{c}{ $t_v=4$}
\\ \cline{2-10}

  & FSTA & FSTG & FDM & FSTA & FSTG & FDM & FSTA & FSTG & FDM \\ \hline\hline
GBM & 1.3732 & 1.3733 &1.3729 & 1.1368 & 1.1369 &1.1364& 0.8429 & 0.8428 &0.8429\\
Merton  & 1.4817 & 1.4814& 1.4800& 1.2261 & 1.2259 &  1.2260& 0.9085 &0.9083 &0.9082 \\
Kou  & 1.4565 & 1.4562 & 1.4556 & 1.2054 & 1.2052 &  1.2064 &0.8934 &0.8933  &0.8942\\
VG & 1.5670& 1.5669 &1.5680 &1.3073 & 1.3069  & 1.3068&  0.9765 & 0.9754 &0.9758  \\
CGMY & 1.8402  & 1.8398 & 1.8402 & 1.5249 & 1.5247  &1.5245& 1.1299 & 1.1297 & 1.1296 \\ \hline
\end{tabular}
 \caption{\small{ESO cost comparison under affine job termination intensity}. For each vesting period $(t_v = 0, 2, 4)$, the three columns  contain ESO costs  computed, respectively, by the two FST methods  in Section \ref{sect-affine} and  Remark  \ref {sect-general} and a  finite difference method. The job termination intensities are $\lambda(x) = \tilde \lambda(x) = -0.02 x+ 0.2$, while other parameters are the same as in Table \ref{Tab_constant}.}\label{Tab_costaffine}
\end{table}

\newpage

  \section{Risk Analysis for ESOs}  \label{sec-5}
Over the life of an ESO, the option value will fluctuate depending on the company stock price movement. From the firm's perspective, an increase in the ESO value implies a higher expected cost of compensation as compared to the initially reported value. For the purposes of risk management and financial reporting, it is important to consider the probability that the ESO cost will exceed a given level.  

In addition, the ESO can   be terminated  early voluntarily or  involuntarily prior to the expiration date.  This also motivates the study of the contract termination probability, which can be a tool to calibrate the job termination  intensity parameter. Such a calibration  would provide a crucial input to the ESO valuation model, and permit the pricing to be consistent with the firm's characteristic. Moreover, it is also interesting to examine the impact of job termination risk  on the contract termination probability.

Generally speaking, the job termination rate can differ under $\mathbb{P}$ and $\mathbb{Q}$. However,  since ESOs are not traded,  market prices are unavailable for inferring the $\mathbb{Q}$ intensity. For this reason and notational simplicity,   we assume  that the historical and risk-neutral job termination rates are  identical.

\subsection{Cost Exceedance Probability}\label{sect-lossp}
Let us evaluate at the current time $t$ the probability  that the ESO cost will exceed a given  level $\ell$ on  a fixed  future date $\tilde T$. To this end, we need to  consider three   scenarios separately. \\

\noindent \textbf{Case 1: $t<\tilde T \le t_v$}\\
  In this case,  the time interval $[t, \tilde{T}]$  lies within  the vesting period $[0,  t_v]$, where the unvested ESO can be forfeited upon job termination. Therefore, the  probability that the ESO value will exceed a pre-specified  level $\ell>0$ at time $\tilde{T}$  is given by $\P_{t,x}\{\tilde C(\tilde T,X_{\tilde T})>\ell\}$, where $\P_{t,x}$ is the historical probability measure with  $X_t =x$. Since  the ESO becomes worthless if  $\tau^{\tilde \lambda}$ arrives before  $\tilde T$, we have
\begin{align}
\P_{t,x}\{\tilde C(\tilde T,X_{\tilde T})>\ell\}&= \P_{t,x}\{\tilde C(\tilde T,X_{\tilde T})>\ell, \tau^{\tilde \lambda} \ge  \tilde T\} + \P_{t,x}\{\tilde C(\tilde T,X_{\tilde T})>\ell, \tau^{\tilde \lambda} <  \tilde T\}\\
&=\P_{t,x}\{\tilde C(\tilde T,X_{\tilde T})>\ell \,|\, \tau^{\tilde \lambda} \ge  \tilde T\}\P_{t,x}\{\tau^{\tilde \lambda} \ge  \tilde T\}. \label{ppp}
\end{align}
In \eqref{ppp}, we notice that $\P_{t,x}\{\tau^{\tilde \lambda} \ge \tilde T\} =e^{-\tilde \lambda (\tilde T-t)}$ since  $\tau^{\tilde \lambda} \sim \exp(\tilde \lambda)$.  Hence,  the evaluation  of  $\P_{t,x}\{\tilde C(\tilde T,X_{\tilde T})>\ell\}$ amounts to computing the conditional probability  $ \P_{t,x}\{\tilde C(\tilde T,X_{\tilde T})>\ell \,|\, \tau^{\tilde \lambda} \ge  \tilde T\}$.

Since $\tilde{C}(\tilde{T},x)$ is  increasing in  $x$,  we can find the  critical log-price  $\bar{x}$  such that $\tilde C(\tilde T,\bar{x})= \ell$ and write  \begin{equation}\label{ptx1} p(t,x):=\P_{t,x}\{\tilde C(\tilde T,X_{\tilde T})>\ell\,|\, \tau^{\tilde \lambda} \ge \tilde T \}=\P_{t,x}\{X_{\tilde T}>\bar{x}\,|\,\tau^{\tilde \lambda} \ge \tilde T \}.\end{equation}
The probability function $p(t,x)$ satisfies the   PIDE problem
\begin{align}
(\partial _{t}+ {\mathcal{L}})p&=0,\qquad \quad ~~(t,x) \in [0, \tilde T] \times \R, \label{ptx}\\
p(\tilde T,x)&=\indic{x>\bar{x}}, \quad ~\,x\in \mathbb R,
\end{align}
where ${\mathcal{L}}$ is the  infinitesimal generator of  $X$ under $\P$.
We observe that  the PIDE \eqref{ptx} for $p(t,x)$ is very similar to \eqref{PIDE of vested ESO} without the inhomogeneous term. Applying  the Fourier transform arguments  from Section \ref{sec-4} yields that
\begin{align}
 p(t,x)=\mathcal F^{-1}[\mathcal F[\indic{x>\bar{x}}](\omega)e^{\Psi(\omega)(\tilde T-t)}](x), \quad t \le \tilde{T}\le t_v.
\label{PIDEProb}
\end{align}
The Fourier transform and its inversion in \eqref{PIDEProb}  can be numerically evaluated by the FFT algorithm. In contrast to the ESO cost, this probability does not involve early exercise and can be computed in one time step.

\begin{remark}\label{remark_lord}
An equivalent  way to obtain \eqref{PIDEProb} is to adapt the convolution method used by \cite{Fang2008}.     To verify this, we express \eqref{ptx1} as
\begin{align}
  p(t,x)&=\int _{-\infty}^{\infty} \indic{{z>\bar{x}}}f_{X_{\tilde T}\,|\,X_{t}}(z)dz
=\int ^{+ \infty}_{-\infty}\indic{x+y>\bar{x}}f_{X_{\tilde T-t}}(y)dy,\label{tildept}
\end{align}
where  $f_{X_{\tilde T}\,|\,X_{t}}(z)$ of $X_{\tilde T}$  is the conditional probability distribution function given $X_{t}=x$. Then, we express \eqref{tildept} in terms of  Fourier transform:
\begin{equation*}
\begin{split}
\mathcal F[  p](t,\omega)&=\int _{-\infty}^{\infty} e^{-i\omega x}\big(\int ^{+ \infty}_{-\infty}\indic{x+y>\bar{x}}f_{X_{\tilde T-t}}(y)dy\big)dx\\
&=\int _{-\infty}^{\infty} e^{-i\omega (u-y)}\big(\int ^{+ \infty}_{-\infty}\indic{u>\bar{x}}f_{X_{\tilde T-t}}(y)dy\big)du\quad  \qquad (u=x+y)\\
&=\int _{-\infty}^{\infty} e^{-i\omega u}\indic{u>\bar{x}}du \int ^{+ \infty}_{-\infty}e^{i\omega y}f_{X_{\tilde T-t}}(y)dy=\mathcal F[\indic{x>\bar{x}}](\omega)e^{\Psi(\omega)(\tilde T-t)}.
\end{split}
\end{equation*}
Lastly, applying inverse Fourier transform to the last equation yields \eqref{PIDEProb}, and hence the equivalence.
\end{remark}

A slightly different  approach  is  to adapt  the Fourier transform method  by \cite{MadanCarr1999}.
To illustrate, we assume without loss of generality  that $X_t=0$ at a fixed time $t< \tilde{T}$, and define
 \begin{equation}
 \tilde {p}(t,z) := \P_{t,0}\{ X_{\tilde T} >z\,|\, \tau^{\tilde \lambda} \ge \tilde T \}=\int _{z}^{\infty} f_{X_{\tilde T-t}}(y)dy,\label{p123}
 \end{equation}
 where $f_{X_{\tilde T-t}}$ is the   probability density function of $X_{\tilde T -t}$ with $X_0=0$.  The value  $z$ can be considered as the difference between the upper threshold $\bar x$ and the current  value of $X$.

 Applying Fourier transform to both sides of \eqref{p123}, we get
\begin{align}
\mathcal F[\tilde p](t,\omega)=\int _{-\infty}^{\infty} e^{-i\omega z}\big(\int _{z}^{\infty}f_{X_{\tilde T-t}}(y)dy\big)dz
=\int _{-\infty}^{\infty}f_{X_{\tilde T-t}}(y) \big(\int _{-\infty}^{y}  e^{-i\omega z}dz \big)  dy.\notag
\end{align}
We notice that the inner integral
$\int _{-\infty}^{y}  e^{-i\omega z}dz={{e^{-i\omega z}} \over -{i\omega}}\big  | ^{y}_{-\infty}$ does not converge, so we incorporate   a damping factor $e^{az}$, for $a>0$,  and   consider
${\tilde p}_a(t,z)=e^{az}\tilde p(t,z)$. Then, the corresponding Fourier transform is given by
\begin{align}
\mathcal F[{\tilde p}_a](t,\omega) =\int _{-\infty}^{\infty}f_{X_{\tilde T-t}}(y) \int _{-\infty}^{y}  e^{(a-i\omega)z}dz   dy={1 \over{a-i\omega}}\phi_{\tilde T-t}(-\omega-ia),\label{lasteq}
\end{align}
where $\phi_{\tilde T-t}$ is the characteristic function of $X_{\tilde T - t}$. In turn,   inverse Fourier transform to \eqref{lasteq} yields
\begin{align}\label{FFTM}
\tilde p(t,z)={e^{-az} \over{2\pi}}\int _{-\infty}^{\infty}{e^{iz\omega} \phi_{\tilde T-t}(-\omega-ia) \over {a-i\omega}}d\omega={e^{-az} \over{\pi}}\int _{0}^{\infty}{e^{i\omega z} \phi_{\tilde T-t}(-\omega-ia) \over{a-i\omega} } d\omega,
\end{align}
where we have used the fact that the Fourier transform of the real function  $\tilde p(t,z)$   is even in its real part and odd in its imaginary part. Lastly, the integral in \eqref{FFTM} is approximated by the FFT algorithm. We remark that $\tilde p(t,z)$ is independent of the choice of $a$, and we choose $a \in [1.5, 2.0]$ for numerical implementation.

In Table \ref{tab_prob}, we present the numerical results for   the ESO cost exceedance  probabilities   under different thresholds $\ell$. The FST-FST method first computes  the ESO cost vector via FST (see \eqref{unvested}) which gives the critical value $\bar x$ (see \ref{ptx1}). In the second step, it applies FST to solve  for the exceedance probability as shown in  \eqref{PIDEProb}. The FFT-FST method differs from the FST-FST method in its second step where  FFT (see \eqref{FFTM}) is used to compute the probability. For the GBM, Merton, and Kou models, the reference values are given by closed-form formulas (see Appendix B.2).  The numerical results show that  both Fourier transform methods are very accurate.\\

\begin{table}[h!]
\centering
\begin{tabular}{c|ccc|ccc}
\hline
 & \multicolumn{3}{c|}{$\ell = 1.1$ $\tilde C(0,0)$} & \multicolumn{3}{c}{$\ell = 1.2$ $\tilde C(0,0)$}\\

Model & Reference Value & FST-FST &  FFT-FST   & Reference Value &  FST-FST &  FFT-FST
\\ \hline
\hline
GBM  &0.2534 & 0.2532  & 0.2535 &  0.0868 &  0.0869 & 0.0869  \\
Merton &0.2607 & 0.2608 & 0.2607  &  0.0943 & 0.0942 & 0.0943 \\
Kou  & 0.2431 &   0.2430 & 0.2431 & 0.0799 & 0.0797 & 0.0798 \\ \hline
\end{tabular}

 \caption{\small{Cost exceedance probability in case 1. Here, $t=0, \tilde T =10/252, t_v=2, T=8$ and other parameters are the same as in Table \ref{Tab_constant}. }}\label{tab_prob}
\end{table}

\noindent  \textbf{Case 2: $t_v < t< \tilde T \le T$}\\
After  the vesting period, the employee intends to exercise at any $\tau ^* \le T$, but may be forced to exercise at $\tau ^{\lambda}$. The probability of interest is
 \begin{align}
\hat p(t,x):= \P _{t,x} \{  C(\tilde \tau ^* \wedge \tau ^{\lambda}, X_{\tilde \tau ^* \wedge \tau ^{\lambda}} ) \ge \ell \},
\end{align}with $\tilde \tau ^{*} = \tau ^{*} \wedge \tilde T$.  To compute this, we first identify the critical   log-price $\bar{x}(t)$, such that $C(t,\bar{x}(t))=\ell$, for $t\in [0, \tilde T]$. In turn, we write down the corresponding  inhomogeneous PIDE
\begin{align}
(\partial _t+ {\mathcal{L}})\hat p-\lambda \hat p+\lambda\1_{ \{ x \ge \bar{x}(t) \}} =0,\label{case2}
\end{align}
for $(t,x) \in (t_v, \tilde T) \times \R$. The boundary and terminal  conditions depend on the relative positions of the critical log-price $\bar{x}(t)$ and  optimal exercise boundary  $x^*(t) :=\log({{s^*(t)}/S_0} )$. Precisely,   if $\bar{x}(t) \le x^*(t)$, then we set  $\hat p(t,x)=1$ at time $t$ for $x \ge x^*(t)$. If $\bar{x}(t) > x^*(t)$, we set $\hat p(t,x)=1$ for $ x \ge \bar{x}(t)$, and $\hat p(t,x)=0$  for $x^*(t) < \bar{x}(t)$. As for  the terminal condition, if $\bar{x}(\tilde T) \le K$,  then we set $\hat p(\tilde T,x)= 1$, for $x \ge K$. If $ \bar{x}(\tilde T) > K $, then we have $\hat p(t,x)=1$ for $ x \ge \bar{x}(t)$,  and  $\hat p(t,x)=0$ for $  K \le x < \bar{x}(t)$.

For numerical implementation, we first solve for the ESO values $C(t,x)$, which are then used to  determine  the critical log-price $\bar{x}(t)$, and the associated optimal exercise boundary $s^*(t)$ that gives $x^*(t)$.  With these, we apply Fourier transform   implicit-explicit method, as discussed in Section \ref{sect-general}, to the PIDE problem \eqref{case2}. Iterating backward in time, we have for each time step
\begin{align}
&\hat p(t_{m-1},x)\notag\\
&=\mathcal{F}^{-1}\big[\mathcal{F}[\hat p(t_m,x)](\omega)e^{(\Psi(\omega)-\lambda)(t_m -t_{m-1})}
+{\mathcal{F}[\1_{ \{ x \ge \bar{x}(t_{m-1}) \}}](\omega) \over \Psi(\omega)-\lambda}(e^{(\Psi(\omega)-\lambda)(t_m -t_{m-1})}-1)\big](x),
\end{align}
for $m=M, \ldots, 1,0$, along with the  boundary conditions. In  Figure \ref{Lmaturity},  the cost exceedance probability rises  as the horizon $\tilde T$ lengthens. The probability also increases as the job termination intensity $\lambda$ decreases.\\

\begin{figure}[thb]
\begin{center}
\includegraphics[width=4.5in, angle=0]{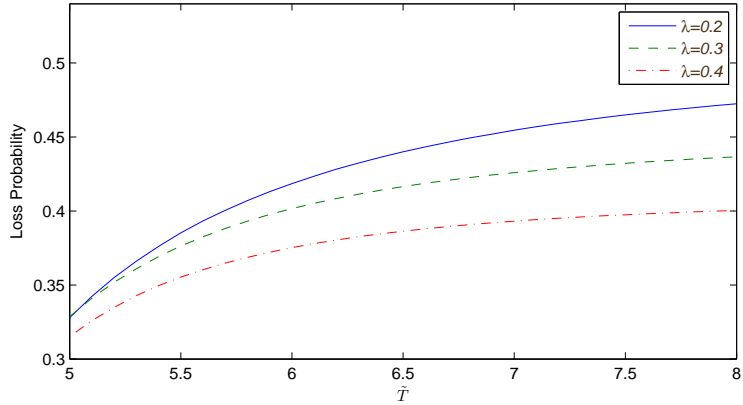}
\caption{ ESO cost exceedance  probability in case 2 increases with $\tilde T$, and decreases with job termination rate $\lambda$. This example is based on the Kou  model with common parameters: $S_4=K=10, r=0.05, p=0.5, q=0.04, \mu=0.08,\tilde \lambda =0.1, T=8, t=4, \sigma=0.2,\alpha =3,\eta_+=50, \eta_-=25$}.\label{Lmaturity}
\end{center}
\end{figure}

\noindent  \textbf{Case 3: $t \le t_v < \tilde T \le T$}\\
This scenario is  a combination of case 1 and case 2. The  ESO is  forfeited if $\tau ^{\lambda} < t_v$, as in case 1. However, if that does not happen, then case 2 applies after the vesting. Consequently, we consider the following probability for ESO cost exceedance
\begin{align}
\bar{p}(t,x):=\E ^{\P}  _{t,x} \{ \1_{\{ \tau ^{\tilde \lambda} \ge t_v\} } \hat p(t_v,X_{t_v})    \}, \quad 0 \le t \le t_v,
\end{align}
if the job termination has not occurred by time $t$. The corresponding PIDE problem is
\begin{align}
(\partial _t+{\mathcal{L}})\bar{p}-\tilde \lambda \bar{p} &=0, \qquad\qquad~~ (t,x)\in [0,t_v)\times \R,\notag\\
\bar{p}(t_v,x)&=\hat p(t_v,x), \qquad x \in \mathbb R.
\end{align}The  solution via Fourier transform is given by
\begin{align}\bar{p}(t_v,x)&= \hat p(t_v,x),\label{tp1}\\
\bar{p}(t,x)&=\mathcal{F}^{-1}\big [\mathcal{F}[\hat p(t_v,x)](\omega)e^{(\Psi(\omega)-\tilde \lambda)(t_v-t)}\big ](x),\quad \text{ for } ~t\le t_v.\label{tp2}
\end{align} Here, the probability  $\hat p(t_v,x)$  from case 2 is used as the input, and  then $\bar{p}(t,x)$ can be computed directly without recursion by \eqref{tp2} for any time $t\le t_v$.

\subsection{Contract Termination Probability}
 The ESO contract can be terminated either due to job termination before   vesting, or  voluntary/forced exercise after   vesting. We now study the ESO contract termination probability during a given time interval $[t,\tilde T]$.  Again, we consider  three different scenarios. \\

\noindent  \textbf{Case 1: $ t<\tilde T \le t_v$ }\\
During the vesting period, the contract termination is totally due to job termination before $t_v$, of which the probability  is simply  $1-e^{-\tilde \lambda (\tilde T-t) }$. \\
   \\
 \noindent \textbf{Case 2: $t_v \le t< \tilde T \le T$ }\\
After vesting,  contract termination can arise from   either  involuntary exercise due to  job termination, or the holder's voluntary exercise. For calculation purpose, we divide the contract termination probability into two parts according to whether job termination occurs before or after $\tilde T$.

First, we consider the scenario where  job termination does not occur during  $[t, \tilde T]$ and  the ESO holder  voluntarily exercises the ESO. This corresponds to the  probability \begin{align}\label{pptt}
 \P  _{t,x}\{ \tau ^* \le \tilde T,  \tau ^{\lambda} > \tilde T \} = \P  _{t,x}\{ \tau ^* \le \tilde T\}  e^ {- \lambda (\tilde T - t)},
\end{align}
 where $\tau ^*$ is the holder's  optimal exercise time. To evaluate \eqref{pptt},  we solve for $ \hat h(t,x): = \P  _{t,x}\{ \tau ^* \le \tilde T\}$ from  the PIDE
\begin{align}
(\partial _{t}+{\mathcal{L}})\hat h=0,
\end{align}
for $(t,x) \in (t_v, \tilde T) \times \R$, with the boundary condition $\hat h(t,x)=1$, for $x > x^*(t)$, and terminal condition $\hat h(\tilde T,x)=\1_{\{x \ge x^*(\tilde T)\}}$, where $x^*(t) :=\log({{s^*(t)}/{S_0}} )$. For numerical solution, we apply  recursively
\begin{align}
\begin{cases}
{
\hat h(t_{m-1},x)=\mathcal F^{-1}[\mathcal F[\hat h(t_{m},x)](\omega)e^{\Psi(\omega)(t_m - t_{m-1})}](x),  \quad x \le x^*(t) }\\
{\hat h(t_{m-1},x)=1, \quad x > x^*(t). }
\end{cases}
\end{align}

The other  scenario is when   job termination arrives before $\tilde T$. Consequently, the total contract termination probability $h(t,x)$ is the sum
 \begin{align}\label{hhh}h(t,x) := e^ {- \lambda (\tilde T - t)} \hat h(t,x)+ 1 - e^{-\lambda (\tilde T - t)} .\end{align}
  From this expression, we observe that the contract termination probability is increasing with job termination intensity $\lambda$ since  the optimal exercise boundary is decreasing with $\lambda$, and so is $(1- \hat h(t,x))$.

Alternatively, we  look at  the employee's  voluntary exercise probability  \[h^v(t,x) := \P  _{t,x}\{ \tau ^* < \tau ^{ \lambda} \wedge \tilde T \}.\] This probability satisfies the PIDE
\begin{align}
(\partial _{t}+{\mathcal{L}})h^v - \lambda  h^v=0,\label{pidehv}
\end{align}
for $(t,x) \in (t_v, \tilde T) \times \R$, with the boundary condition $ h^v(t,x)=1$, for $x > x^*(t)$, and terminal condition $ h^v(\tilde T,x)=\1_{\{x \ge x^*(\tilde T)\}}$, where $x^*(t) :=\log({{s^*(t)} / {S_0}} )$. The numerical solution is found from backward iteration with
\begin{align}
\begin{cases}
{
h^v(t_{m-1},x)=\mathcal F^{-1}[\mathcal F[h^v(t_{m},x)](\omega)e^{(\Psi(\omega)-\lambda)(t_m - t_{m-1})}](x),  \quad x \le x^*(t) }\\
{h^v(t_{m-1},x)=1, \quad x > x^*(t), }
\end{cases}
\end{align}for $m=M, \ldots, 1,0$.

Figure \ref{fig_last} illustrates  the  contract termination probability on top of the  voluntary exercise probability.  From PIDE \eqref{pidehv}, we observe two competing factors governing  the effect of job termination intensity $\lambda$ on the  voluntary exercise probability. On one hand, a higher job termination intensity $\lambda$ implies a lower optimal exercise boundary, which in turn  increases the voluntary exercise probability. On the other hand, a higher job termination intensity will more likely force early exercise before the stock price reaching the holder's optimal exercise boundary. This reduces  the voluntary exercise probability.  In  Figure \ref{fig_last} (left), we  see   that voluntary exercise probability is decreasing with post-vesting job termination intensity $\lambda$, so in this case the job termination effect outweighs  the effect of lowered  exercise boundary, and therefore, reduces the probability of voluntary exercise. Also, in Figure \ref{fig_last} (right),  the contract termination probability is increasing with job termination intensity $\lambda$, as expected.\\

\begin{figure}[h!]
    \includegraphics[width=3.5in]{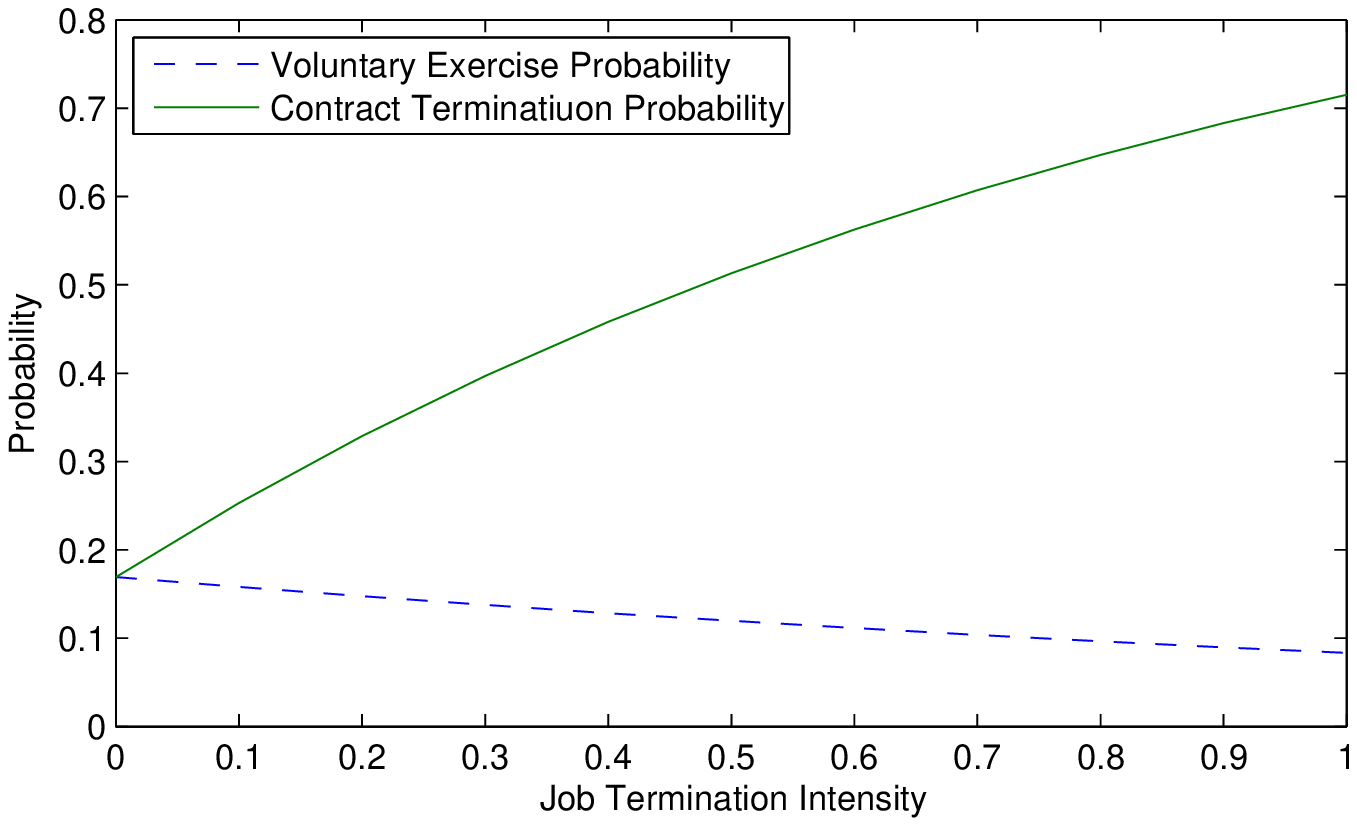}
    \includegraphics[width=3.5in]{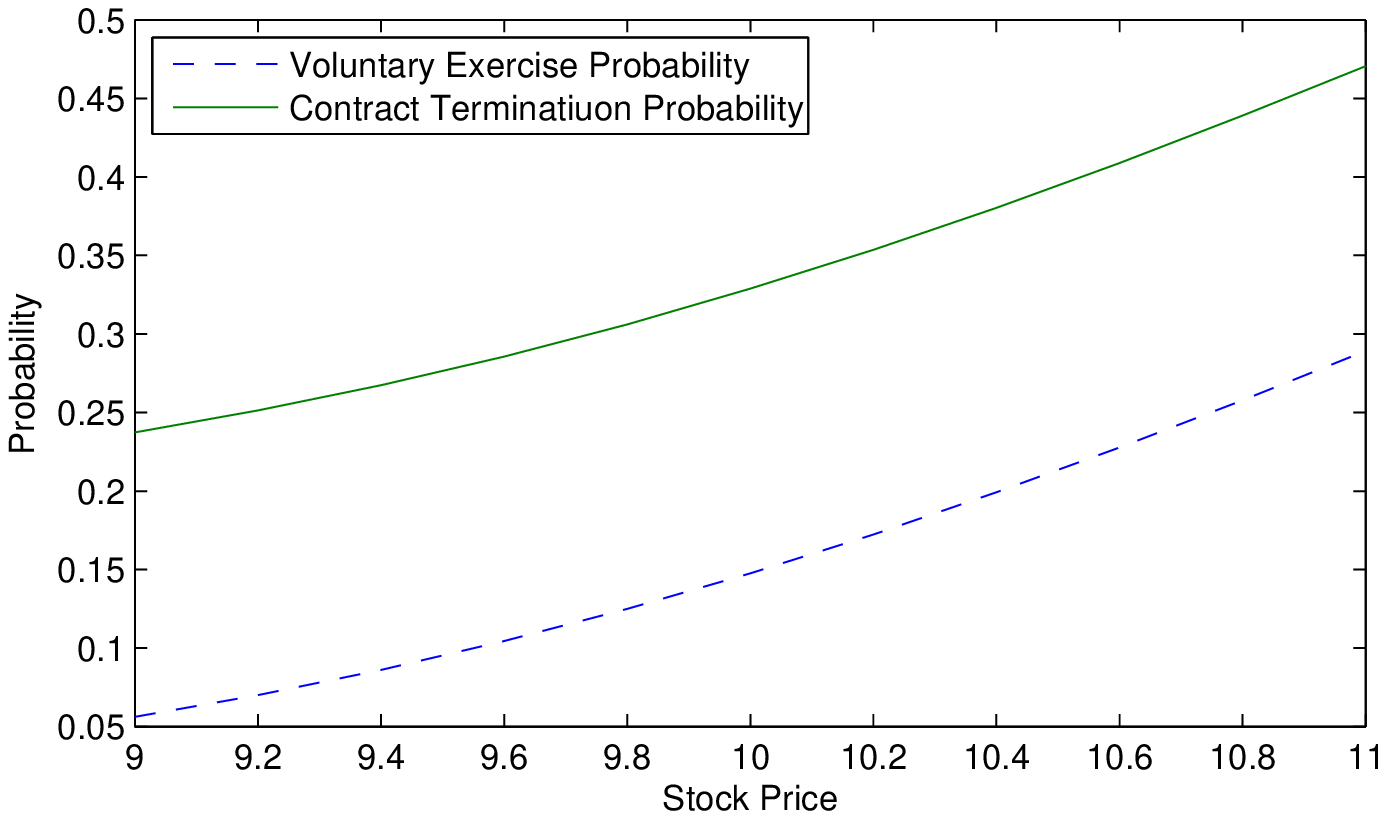}
         \caption{(Left) Job termination intensity increases contract termination probability, but decreases voluntary exercise probability. (Right) Stock price increases both contract termination probability and voluntary exercise probability with $\tilde \lambda =0.1$ and $\lambda =0.2$. This example is based on the GBM model with common parameters: $S_6=10, K=9, r=0.05, \sigma=0.2, q=0.04, \mu=0.08, t_v=2, T=8, t=6, \tilde T=7$.}\label{fig_last}
\end{figure}

\noindent  \textbf{Case 3: $t \le t_v<\tilde T \le T$}\\
This scenario is a combination  of cases 1 and  2 above. The contract termination can occur before or after vesting. During  $[t,t_v]$, only job termination can cancel the contract. If there is no job termination before  $t_v$, then the contract termination resembles that in case 2. Therefore, the contract termination probability is the sum
\begin{align}
1-e^{-\tilde \lambda (t_v -t)}+\underbrace{\E^{\P}_{t,x}\{ h(t_v,X_{t_v}) \1 _{\{ \tau ^{\tilde \lambda} \ge t_v \} }\}}_{=:\tilde{h}(t,x)},
\end{align}
where $h(t,x)$ is given in \eqref{hhh}. Hence,   $\tilde h(t,x)$ satisfies, for $(t,x) \in [0,t_v) \times \R$, the PIDE
\begin{align}
(\partial _{t}+{\mathcal{L}})\tilde h -\tilde \lambda \tilde h&=0, \qquad\qquad\quad~\,
\end{align}At time $t_v$, we set $\tilde h(t_v,x)= h(t_v,x)$, where $h(t_v,x)$ is computed   from case 2.
 Again, the FST method discussed in Section \ref{sec-4} can be applied to  solve for $\tilde h(t,x) $.  At any time $t < t_v$, the probability can be  computed in one step (without time iteration) via
\begin{align} \tilde h(t,x)=\mathcal{F}^{-1}\big [\mathcal{F}[h(t_v,x)](\omega)e^{(\Psi(\omega)-\tilde \lambda)(t_v-t)}\big ](x).
\end{align}

%
%
%

\newpage
\section{Perpetual ESO with Closed-Form Solution}\label{sec-perp}
We now discuss the valuation   of a perpetual  ESO  under the GBM model. In contrast to the   valuation under the general \lv framework, the perpetual ESO  admits a closed-form solution, and thus represents a highly  tractable alternative.    We first consider a vested  perpetual ESO, whose value can be expressed  in terms of an optimal stopping problem, namely,
\begin{align}
V(s) &=\sup_{ \tau \in \setT_{0,\infty}}  \E^{\Q}  \biggl\{\, e^{-  (r+\lambda ){\tau}} ( S_{\tau }
-K)^+ + \int_{0}^{\tau} e^{- (r+\lambda )u } \lambda  ( S_u -K)^+
du \,|\,S_0=s\biggr\},\label{Cost_infinity}
\end{align}
where $S_t$ is the real stock price.   The associated variational inequality is
\begin{align}
\min\biggl \{-\hat {\mathcal{L}}V+(r+\lambda)V-\lambda (s-K)^+,~ V(s)-(s-K)^+\biggl \}=0, \quad s\in \R_+.
\label{VIinf}
\end{align}

A similar inhomogeneous variational inequality has been derived and solved  for  the problem   of pricing American puts with   maturity randmization (Canadization) introduced by   \cite{Carr_1998}.  Indeed, the   perpetual ESO  can be considered as an American call whose maturity is an exponential random variable.  For a mathematical analysis  on the  Canadization   of American options with a \lv underlying, we refer to \cite{kypri_Canada}.    Next, we present the closed-form solution for this ESO valuation problem.
 
\begin{proposition} \label{thmV}Under the  GBM model,  the value of a vested  perpetual ESO  is given by \begin{align}\label{formulac}
V(s)= \begin{cases}
Ds^{\gamma_+} \quad & \text{if} \,\, s < K,\\
As^{\gamma_+} + B s^{\gamma_-} +  { {\lambda} \over {\lambda +q}} s  -{ {\lambda} \over {r+\lambda} } K \quad &\text{if} \,\, K \le s < s^*,\\
s-K \quad & \text{if}\,\, s \ge s^*,
\end{cases}
\end{align}
where
\begin{align}
\gamma_\pm &={ {  (q-r+ {\sigma^2 \over 2} ) \pm \sqrt{ (q-r+ {\sigma ^2 \over 2} )^2 + 2(r+\lambda ) \sigma ^2  }          }   \over {\sigma ^2}  },\label{dpm1}\\
B&=  {  {   \lambda  (   1-\gamma_+ ) }  \over   { \gamma_- (\gamma_+ - \gamma_-)(\lambda +q)  }  } K^{1-\gamma_-},\label{dpm2}\\
  A &=    {q \over {(\lambda +q) \gamma_+}  }   \,(s^{*})^{1-\gamma_+} -     { {\gamma_-} \over {\gamma_+} }  B  \,(s^*)^{ \gamma_--\gamma_+},\label{dpm3}\\
D& =  A+ B \, K ^{\gamma_- -\gamma_+} +  \frac{ \lambda (r - q)}{ (\lambda+q)( \lambda + r)}\,K^{1-\gamma_+}\label{dpm4}.
\end{align} The optimal exercise threshold $s^*\in(K,\infty)$ is   uniquely determined from
\begin{align}
B (1 - {  \gamma_- \over {  \gamma_+} }) (s^*)^{\gamma_-} - (1- { 1 \over {\gamma_+} })  ( { {q} \over { \lambda +q} }) s^*+ { {r} \over {r+\lambda} } K  = 0 .\label{S*}
\end{align}
\end{proposition}

%
%
%

A number of interesting observations can be drawn from  Theorem \eqref{thmV}. First, in the case without job termination ($\lambda =0$), we have $B=0$ and   $A=D$. This implies that the perpetual  ESO reduces to  an ordinary American call with the well-known price formula
\begin{align}
V(s)= \begin{cases}
Ds^{\gamma_+} \quad \text{if} ~\, s < s^*,\\
s-K \quad \text{if}~\, s \ge s^*,
\end{cases}
\end{align}  where  $s^*= \frac{ \gamma_+}{ 1-   \gamma_+}K$.    This result dates back to   \cite{Samuelson65} and \cite{McKean1965}, and is also used in the real option literature (see, for example, \cite{McDonald86}). 
 Furthermore, if  both $\lambda$ and $q$ are zero, then  we can  see that $s^*= \infty$, which means that it is optimal for the option holder to never exercise the option. This is expected since the ESO now resembles an ordinary American call without dividend.


\begin{figure}[h!]
\begin{center}  \includegraphics[width=5in]{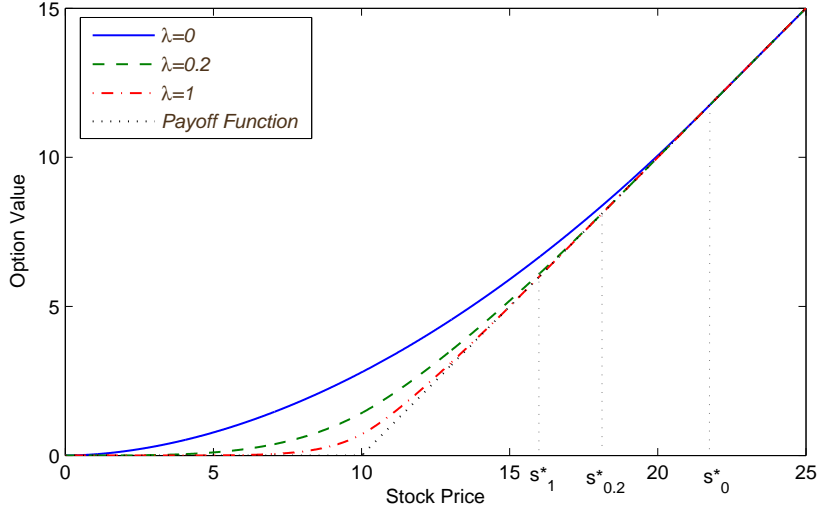}
   \caption{\small {As job termination rate increases ($\lambda \in \{0, 0.2, 1\}$),  the  vested ESO cost decreases  and  the corresponding optimal exercise threshold is lowered  ($s^*_0> s^*_{0.2}> s^*_{1}$). Parameters: $K=10, r=0.05, \sigma=0.2, q=0.04$.} }\label{fig_Perpetual}
   \end{center}
\end{figure}
%
%

When a  vesting period of $t_v$ years is imposed, we compute the ESO cost  at time $t$ from the conditional expectation
\begin{align}\tilde V(t,s) = \E ^{\Q}   \{   e^{-(r+\lambda) (t_v-t)} V(S_{t_v}) \,|\,S_t = s \}.\label{unvestedperpeso}\end{align}
 By substituting   the vested ESO cost function  $V(s)$  into  \eqref{unvestedperpeso}, and recognizing   that $S^{a}_{t_v}$, for any $a\in\R$, is  lognormal, we can directly compute the unvested ESO cost.

\begin{corollary}Under the GBM model,  the unvested perpetual ESO cost  admits the formula
\begin{align}
\tilde V(t,s) &= e^{-(r+\lambda)(t_v-t)} \bigg\{ D s^{\gamma_+} e^{(m_1 + {{\sigma_1^2}  \over 2})} \bar\Phi\bigg({ {\gamma_+ \ln (s/K) + m_1 + \sigma_1^2}  \over {\sigma_1}} \bigg)\label{formulaunvestperp}\\
&~+ A s^{\gamma_+} e^{(m_1 +{{\sigma_1^2}  \over 2})} \bigg[\Phi\bigg( { {\gamma_+ \ln (s/K) + m_1  +\sigma_1^2}  \over {\sigma_1}} \bigg) - \Phi\bigg( { {\gamma_+ \ln (s/s^*) + m_1 +\sigma_1^2}  \over {\sigma_1}}  \bigg)   \bigg]\notag\\
&~+  B s^{\gamma_-} e^{(m_2+{{\sigma_2^2}  \over 2})} \bigg[ \Phi\bigg( { {\gamma_- \ln(s/s^*) + m_2 +\sigma_2^2}  \over {\sigma_2}} \bigg) -\Phi\bigg( { {\gamma_- \ln(s/K) + m_2  +\sigma_2^2}  \over {\sigma_2}} \bigg) \bigg]\notag\\
&~+s(1+ { {\lambda} \over {\lambda +q}}) e^{(m_3+ {{\sigma_3^2}  \over 2})} \Phi\bigg( { {\ln(s/K) + m_3  +\sigma_3^2}  \over {\sigma_3}} \bigg)- (1+{ {\lambda} \over {r+\lambda} }) K \Phi\bigg( { {\ln (s/K)+ m_3 +\sigma_3^2}  \over {\sigma_3}} \bigg) \bigg\},\notag
\end{align}where  $\bar \Phi$ is the standard normal complementary  c.d.f. and
 \begin{align}
m_1&= (r-\frac{\sigma^2}{2})(t_v-t)\, \gamma_+, \quad \sigma_1 =  \sigma \sqrt{t_v-t} \,\gamma_+,\\
m_2 &= (r-\frac{\sigma^2}{2})(t_v-t)\, \gamma_-,  \quad \sigma_2 =  -\sigma \sqrt{t_v-t}\, \gamma_-,\\
   m_3 &= (r-\frac{\sigma^2}{2})(t_v-t),    \,\qquad \sigma_3 =\sigma \sqrt{t_v-t}.
 \end{align}
\end{corollary}

In contrast to its vested counterpart, the unvested ESO cost is time-dependent. In Figure \ref{fig_Perpetual2} (left), we observe that the perpetual   ESO cost  decreases as the vesting period increases. A   higher job termination rate not only decreases the ESO cost (see Figure \ref{fig_Perpetual}) but also the exercise threshold $s^*$ (see Figure \ref{fig_Perpetual2} (right).  

\begin{figure}[h!]
\begin{center}  \includegraphics[width=3.15in]{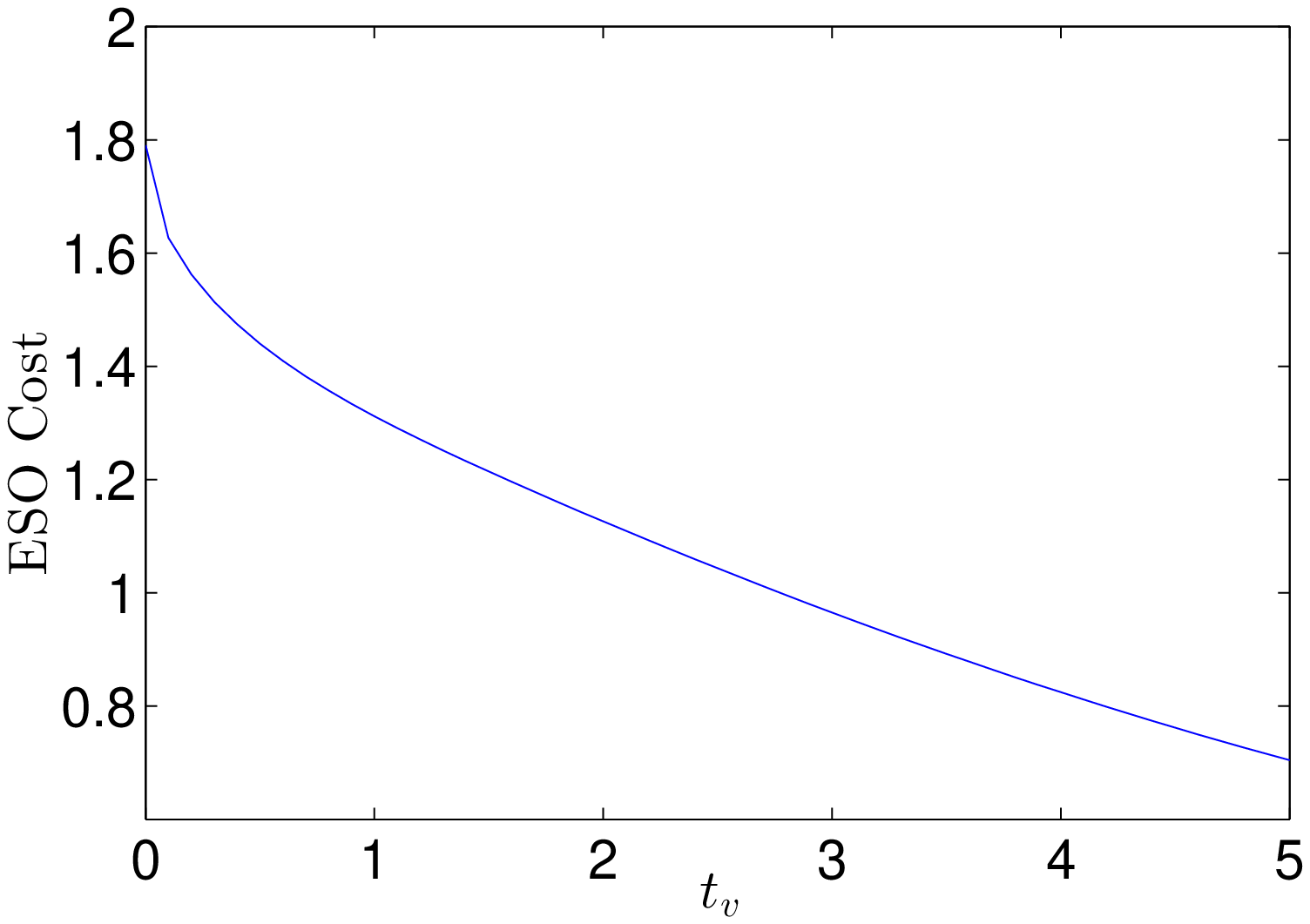}~
    \includegraphics[width=3in]{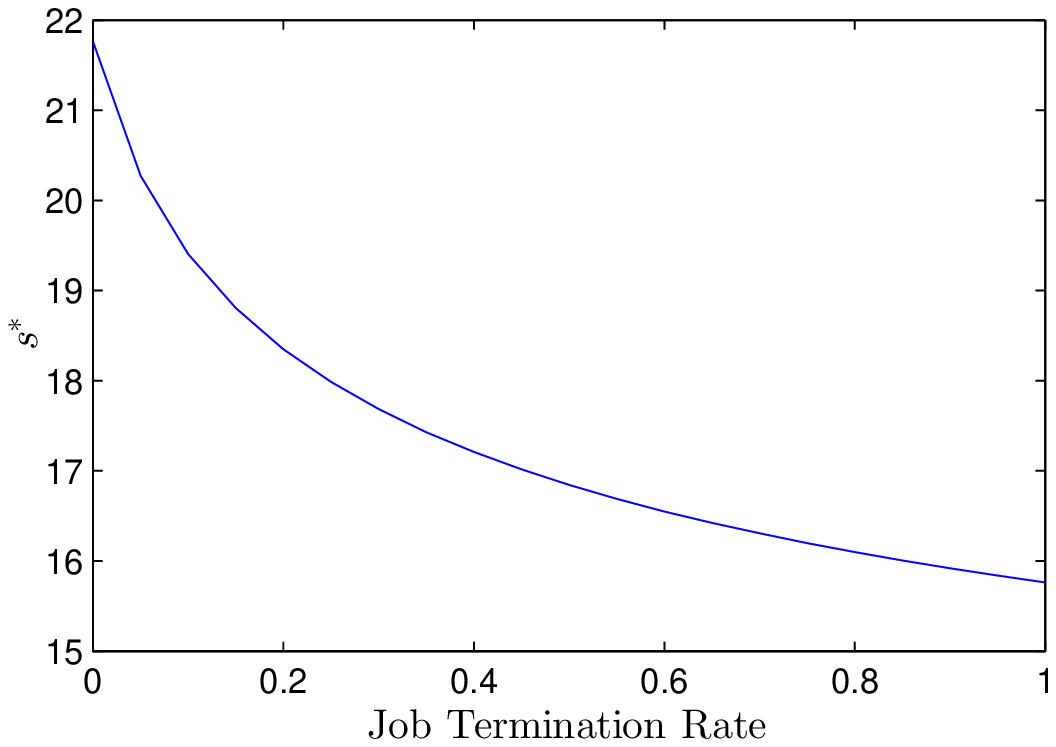}
   \caption{\small {The perpetual ESO cost decreases  as vesting period increases (left) and as job termination rate  increases (right). Parameters: $S_0=K=10, r=0.05, \sigma=0.2, q=0.04, t=0$ (left), $t_v=0$ (right).} }\label{fig_Perpetual2}
   \end{center}
\end{figure}

\section{Conclusions}\label{sec-6}
We have provided  the analytical and numerical studies for the valuation and risk analysis of ESOs under \lv price dynamics.  Our results are useful for reporting  ESO cost, as mandated by regulators,   and for understanding holder's exercise behavior.   In particular, we show job termination risk has a direct effect on the ESO holder's exercise timing, which in turn affects the ESO cost as well as contract termination probability.  For  future research, risk estimation  for large ESO portfolios  is both practical and challenging. Other related issues include the incentive effect and optimal design  of ESOs and other compensation schemes, such as restricted stocks. Lastly, our valuation framework can also be applied to pricing American options with liquidity, default, or other event risks. This would require an appropriate modification of the payoff at the exogenous termination time.

\appendix
\section{Appendix}
%
%
%
%

\subsection{Proof of Proposition \ref{proplast} } \label{Appendix1}
Let $C_1(t,x)$ and $C_2(t,x)$ be the vested ESO cost associated with $\lambda _1$ and $\lambda _2$, respectively, and assume $\lambda_1 < \lambda_2$. We define an operator $\mathcal M$ by
\begin{align}
\mathcal M _i C(t,x)= (\partial _t+\hat {\mathcal{L}})C-(r+\lambda_i)C+\lambda_i(S_0e^x-K)^+.
\end{align}
From the variational inequality (\ref{VI}), we   see that $ \mathcal M _i C_i \le 0 $. We choose a point $(t,x)$ in the continuation region of $C_2$, which means that $\mathcal M_2 C_2 =0$. Since $\lambda _1 < \lambda _2$ and $C_2 > (S_0e^x-K)^+$, direct substitution shows that $\mathcal M_1 C_2 >0$.

Next, we define the process
\begin{align}
m(t,X_t)=e^{-(r+\lambda _1)t}C_2(t,X_t)+\int _0 ^t e^{-(r+\lambda _1)u} \lambda _1 (S_0e^{X_u}-K)^+du, \quad t\ge 0.
\end{align}
Using the fact that  $\mathcal M_1 C_2 >0$ and  Optional Sampling Theorem, we deduce that, for any $\tau\in \setT_{t,T}$,
\begin{align}
\E ^{\Q} _{t,x}\big \{ m(\tau, X_{\tau})  \big \} \ge m(t,x).
\end{align}
In particular, we denote $\tau _1 ^*$ and $\tau _2 ^*$ as the optimal stopping times associated with $C_1$ and $C_2$, and get
\begin{align}
C_2(t,x) & \le \E ^{\Q}_{t,x} \biggl\{  e^{-(r+\lambda _1)(\tau _2 ^*-t)}C_2(\tau _2 ^*,X_{\tau _2 ^*})+\int _t ^{\tau _2 ^*} e^{-(r+\lambda _1)(u-t)} \lambda _1 (S_0e^{X_u}-K)^+du \biggl \} \\
&= \E ^{\Q}_{t,x} \biggl\{  e^{-r(\tau _2 ^* \wedge \tau ^{\lambda_1} -t)}(S_0e^{X_{\tau _2 ^* \wedge \tau ^{\lambda_1}}}-K)^+ \biggl \} \\
& \le C_1(t,x).
\end{align}The last inequality follows since $\tau_2^*$ is one candidate stopping time for the optimal stopping  value function $C_1$.  Hence, we conclude that $C_1(t,x) \ge C_2(t,x) > (S_0e^{x}-K)^+$. This implies that any  point $(t,x)$ in the continuation region of  $C_2$ must also lies in  the continuation region of $C_1$,  which means that the optimal exercise boundary for $C_1$ dominates that for $C_2$.

As for the  unvested ESO,  the job termination intensity reduces its  terminal values, and increases the probability of forfeiture (with payoff zero) during the vesting period. As a result, a higher job termination intensity also reduces  the unvested ESO cost. 

\subsection{Proof of Proposition \ref{thmV}}
 We conjecture that it is optimal to exercise the ESO as soon as the stock reaches some  level $s^* >K$.  Then, we split the stock  price domain into three regions: $[s^*, \infty)$, $[K, s^*)$, and $[0, K)$. In region 1, we have $s \ge s^*$ and   $V(s)= s-K$.  In region  2, the ESO cost solves the  inhomogeneous  ODE
\begin{align}\label{ODE11}
{\sigma ^2s^2 \over 2}  { V''(s)  } + (r-q) s V'(s) - (r + \lambda) V(s) +\lambda (s-K)=0.
\end{align}
One can check by substitution that  the general  solution  to \eqref{ODE11} is given by\begin{align}
V(s) = As^{\gamma_+} + B s^{\gamma_-} +  { {\lambda} \over {\lambda +q}} s  -{ {\lambda} \over {r+\lambda} } K,
\end{align}  where $\gamma_-$ and $\gamma_+$ are  given in \eqref{dpm1}.

In region 3, since the option is out of the money, we have the ODE
\begin{align}
{\sigma ^2s^2 \over 2}   V{''}(s) + (r-q) s V{'}(s) - (r + \lambda) V(s)=0,
\end{align}
whose solution is the form $V(s) = Ds^{\gamma_+} + E s^{\gamma_-}$. Since $V(s) \rightarrow 0$ as $s \rightarrow 0$, it follows that  $E=0$.

 To solve for the constants  $A,B,D$, along with  the critical stock price $s^*$,  we   apply continuity and smooth-pasting conditions at $s=K$ and $s=s^*$ to get
\begin{align}
\lim _{s \uparrow K} V(s) = V(K) \quad &\Rightarrow \quad D K^{\gamma_+} = A K^{\gamma_+} + B K^{\gamma_-} + ({ {\lambda} \over {\lambda +q}} -  { {\lambda} \over {r+\lambda} } )K,\\
\lim _{s \uparrow K} V'(s) = V'(K) \quad &\Rightarrow \quad  \gamma_+ D K^{\gamma_+ -1}=   \gamma_+A K^{\gamma_+ -1} + \gamma_-  B K^{\gamma_- -1}+  { {\lambda} \over {\lambda +q}}, \\
\lim _{s \uparrow s^*} V(s) = V(s^*)\quad &\Rightarrow \quad A (s^{*})^{\gamma_+} + B (s^{* })^{\gamma_-} +  { {\lambda} \over {\lambda +q}} s^*  -{ {\lambda} \over {r+\lambda} } K=s^*-K,\label{s1}\\
\lim _{s \uparrow s^*} V'(s) = V'(s^*)  \quad &\Rightarrow  \quad  \gamma_+A  (s^{*}) ^{\gamma_+ -1} +\gamma_-   B (s^{ *}) ^{\gamma_- -1} +  { {\lambda} \over {\lambda +q}}  =1.\label{s2}
\end{align}
Solving  this  system of equations yields \eqref{dpm2}-\eqref{dpm3}. In particular, we see that $B>0$ since $\gamma_+>1$ and  $\gamma_-<0$.

 From \eqref{s1}-\eqref{s2}, the threshold $s^*$ satisfies  the equation
\begin{align}
f(s^*):= B (1 - {  \gamma_- \over {  \gamma_+} }) (s^*)^{\gamma_-} - (1- { 1 \over {\gamma_+} })  ( { {q} \over { \lambda +q} }) s^*   +  { {r} \over {r+\lambda} } K= 0.\label{S*}
\end{align}
 To show this has a unique real solution, we note that $f$ is continuous, and
\begin{align}
f'(s^*) = \gamma_- B (1 - {  \gamma_- \over {  \gamma_+} }) (s^*)^{\gamma_- -1} + (1- { 1 \over {\gamma_+} })  ( { {-q} \over { \lambda +q} }) < 0,
\end{align}
since $B >0, \gamma_-<0$, and $\gamma_+ >1$. In addition, we have the limits: $\lim_ { s^* \downarrow 0} = { {r} \over {r+\lambda} }  >0$ and  $\lim _{s^* \uparrow \infty} = -\infty$, as well as  $f(K) =  { K  \over {\gamma_+} } >0$.  Together, this implies $f(s^*) = 0$ has a unique real root $s^*>K$. Hence,  we  obtain formula \eqref{formulac} for $V$. By direct substitution, it satisfies the  VI \eqref{VIinf}.

\subsection{Finite Difference Method for ESO Valuation}\label{app_fdm}
We summarize the finite difference method (FDM) for computing  the ESO costs in Tables \ref{Tab_constant} and \ref{Tab_costaffine}. For this purpose, we adapt the  FDM algorithm  for European options detailed in \cite{Cont2003} to the current case of early exercisable ESO with job termination risk.

First, we introduce the change of variable $u=T-t$ and denote  $F(u,x)=C(T-t,x)$. Then, the PIDE for the  vested ESO cost in the continuation region becomes
\begin{equation}
{\partial F \over \partial u}=\hat{\mathcal{L}}F-(r+\lambda(T-u,x))F+ \lambda(T-u,x) (S_0e^x-K)^+,
\end{equation}
for $(u,x) \in (0,T-t_v)\times \mathbb R$, with the initial condition $F(0,x)=(S_0e^x-K)^+,\,x\in \mathbb R$, where
\begin{equation}
\hat{\mathcal{L}} F=(r-q){\partial F \over \partial x}+{\sigma^2 \over 2}({\partial ^2 F \over \partial x^2}-{\partial F \over \partial x})+\int^{\infty} _{-\infty}\hat\nu (dy) (F(u,x+y)-F(u,x)-(e^y-1){\partial F \over \partial x}).
\end{equation}

To proceed, we split the operator $\hat{\mathcal{L}} $ into two parts, namely,
\begin{equation}
\hat {\mathcal{L}} F=DF+JF,
\end{equation}
where
\begin{equation}\label{DJ}
DF={\sigma^2 \over 2}{\partial ^2 F \over \partial x^2}-({\sigma^2 \over 2}-r+q+\beta){\partial F \over \partial x}-\alpha F,\quad \text{ and } \quad  JF=\int^{B_r}_{B_l}\hat \nu(dy)F(u,x+y),
\end{equation}
with $\beta=\int^{B_r}_{B_l}\hat\nu(dy)(e^y-1)$.

We define a uniform grid on $[0,T-t_v]\!\times\! [-A,A]$ by $\{(u_n, x_i) : u _n=n \Delta t, n=0,1,2,\ldots,M, x_i= -A+i\Delta x,  i\in {0,1,2 \ldots, N}\}$, with $\Delta t=(T-t_v)/M, \Delta x=2A/N$. In Tables \ref{Tab_constant} and \ref{Tab_costaffine}, we have $M= 4000$ and $N=8000$. Also, denote  ${F_i^n}$ be the cost at the grid point $(u_n, x_i)$.  We use trapezoidal quadrature rule with $\Delta x$ to approximate the integral terms in \eqref{DJ}. To do so, we let $K_l$, $K_r$ be such that $[B_l,B_r] \subset [(K_l-1/2)\Delta x,(K_r+1/2)\Delta x]$, and apply the  approximations
\begin{equation}
\int^{B_r}_{B_l}\hat\nu(dy)F(u,x_i+y)\approx \sum_ {j=K_l} ^ {K_r}\hat\nu _j F_{i+j}, \quad \alpha \approx\sum_ {j=K_l} ^ {K_r}\hat\nu _j, \quad  \beta \approx  \sum_ {j=K_l} ^ {j=K_r}\hat\nu_j {(e^{y_j}-1)} ,\notag
\end{equation}
with
\begin{equation}
  \hat\nu_j=\int^{(j+1/2)\Delta x}_{(j-1/2)\Delta x} \hat\nu(dy).\notag
\end{equation}
The space derivatives are approximated by the  finite differences
\begin{equation}
({\partial  F \over \partial x})_i \approx {{F_{i+1}-F_i} \over \Delta x}, \quad ({\partial ^2 F \over \partial x^2})_i \approx {{F_{i+1}-2c_i+F_{i-1}} \over (\Delta x )^2}.
\end{equation}

Next, we replace $DF$ and $JF$  with their approximations $D _\Delta F$ and  $J _\Delta F$, respectively.  Lastly, we arrive at the following implicit-explicit time-stepping scheme:
\begin{equation}
{{F^{n+1}-F^n}\over \Delta t}=D _\Delta F^ {n+1} +J _\Delta F^ n - \big(r+\lambda(T-(n+1)\Delta t,x)\big)F^ {n+1}+ \lambda \big(T-(n+1)\Delta t,x\big) (S_0e^{x}-K)^+,
\end{equation}
where
\begin{equation}
(D _\Delta F)_ i={\sigma^2 \over 2}{{F_{i+1}-2F_i+F_{i-1}} \over (\Delta x )^2}-({\sigma^2 \over 2}-r+q+ \beta){{F_{i+1}-F_i} \over \Delta x}-\alpha F_i, \quad (J _\Delta F)_ i= \sum_ {j=K_l} ^ {K_r}\hat\nu _j F_{i+j}.
\end{equation}
Due to the early exercise feature, the iteration is coupled with a comparison to the payoff from immediate exercise.  After computing  the vested ESO cost till the end of the vesting period, similar finite difference method can be applied to solve the  PIDE for the unvested ESO cost:
\begin{equation}
{\partial F \over \partial u}=\hat{\mathcal{L}}F - (r+\tilde \lambda(T-u,x))F,
\end{equation}
for $(u, x)\in (0,t_v)\times \mathbb R$.

The above algorithm works for the case when the underlying L\'evy process has  finite activity, with  $\hat\nu(\mathbb R)=\alpha < +\infty$.   In the infinite activity case with  $\hat\nu(\mathbb R)=+\infty$,   we we can  use an auxiliary  process $(X_t ^\varepsilon)_{t\ge 0}$ with the L\a'{e}vy triplet $(\hat{\mu}(\varepsilon),\sigma ^2+\sigma(\varepsilon) ^2,\hat \nu1_{|x|>\varepsilon})$ to approximate the original process $(X_t)_{t\ge 0}$, where
$
\sigma(\varepsilon) ^2=\int_{-\varepsilon} ^{\varepsilon} y^2\hat \nu(dy),
$
and $\hat{\mu}(\varepsilon)$ is again determined by the risk-neutral condition. Therefore, $F^{\varepsilon}$ satisfies
\begin{equation}\label{PIDEcep}
\begin{split}
{\partial F^{\varepsilon} \over \partial u} = \hat {\mathcal{L}}^{\varepsilon}F^{\varepsilon}-(r+\lambda(T-u,x))F^{\varepsilon}+ \lambda(T-u,x)(S_0e^x-K)^+,
\end{split}
\end{equation}
for $(u,x) \in (0,T-t_v)\times \mathbb R$ with   initial condition $F^{\varepsilon}(0,x)=(S_0e^x-K)^+,\,x\in \mathbb R$. Here, the operator $\hat{\mathcal{L}}^{\varepsilon}$ is defined by
 \begin{equation}
\hat{\mathcal{L}}^{\varepsilon}F^{\varepsilon}=({\sigma ^2+\sigma(\varepsilon) ^2 \over 2}) {\partial ^2 F^{\varepsilon} \over \partial x^2}-({\sigma ^2+\sigma(\varepsilon) ^2 \over 2}-r+q+\beta (\varepsilon)){\partial F^{\varepsilon} \over \partial x} -\alpha({\varepsilon})F^{\varepsilon}(x)+\int_{|y| \ge {\varepsilon}}\hat\nu(dy)F^{\varepsilon}(x+y),
\end{equation}
with \[\beta(\varepsilon)=\int_{|y| \ge {\varepsilon}}(e^y-1)\hat\nu(dy) \quad \text{ and } \quad \alpha(\varepsilon)=\int_{|y| \ge {\varepsilon}}\hat\nu(dy).\]
  The PIDE in \eqref{PIDEcep} can be  solved by the same numerical  scheme as in the finite activity case. We apply this finite difference method for comparing with our FST method. For alternative finite difference  methods, especially those designed to address  specific   L\'evy processes, such  as VG and CGMY, we refer to  \cite{MadanHirsa2004},  \cite{Forsyth07}, and references therein.

\subsection{Closed-Form Probabilities}\label{closed form}
The probability $p(t,x)$  that an ESO cost surpasses a given threshold $\bar{x}$, as defined in \eqref{ptx1},  can be viewed as a European digital option with zero interest rate, computed under $\mathbb{P}$. We summarize the corresponding closed-form formulas  under the GBM, Merton, and Kou  models (see Table \ref{lv}).

\noindent (\romannumeral1) Under the GBM model, the ESO cost exceedance probability  is given by
\begin{align}
p(t,x)= e^{-\tilde \lambda (\tilde T -t)}\Phi(\tilde{d}), \quad \text{ where }  \quad \tilde{d}={{ x-\bar{x}  + \mu (\tilde T -t)} \over { \sigma \sqrt{\tilde T -t}}},
\end{align}
and $\Phi$ is the standard normal c.d.f.

\noindent(\romannumeral2) When the company stock price follows the Merton jump diffusion, we have
\begin{align}
p(t,x) = e^{-\tilde \lambda (\tilde T -t)} \sum _{j=0} ^{+\infty} { {e^{ -\alpha (\tilde T -t) } (\alpha (\tilde T -t)) ^j}  \over j!}   \Phi \big({ {x - \bar{x}+ \mu (\tilde T - t) + j \tilde \mu} \over {\sqrt{ \sigma ^2 (\tilde T -t) +j \tilde \sigma ^2 } } } \big).
\end{align}

\noindent(\romannumeral3) In the Kou jump diffusion model, the probability of cost exceedance is given by
\begin{equation}
\begin{split}
p(t,x)=&e^{-\tilde \lambda v} \biggl[{e^{(\sigma \eta_+ )^2 v / 2} \over {\sigma \sqrt{2 \pi v}}  } \sum _{n=1}^{ \infty} \pi _n \sum _{k=1}^{n}P_{n,k} (\sigma \sqrt{v \eta_+} )^k
 \times I_{k-1}(\bar x -x - \mu v; -\eta_+, {{-1} \over {\sigma v} }, -\sigma \eta_+ \sqrt{v} )\\
&+{e^{(\sigma \eta_- )^2v / 2} \over {\sigma \sqrt{2 \pi v}}  } \sum _{n=1}^{ \infty} \pi _n \sum _{k=1}^{n}Q_{n,k} (\sigma \sqrt{v \eta_-} )^k
 \times  I_{k-1}(\bar x -x - \mu v; -\eta_-, {{-1} \over {\sigma v} }, -\sigma \eta_- \sqrt{v} )\\
&+\pi_0 \Phi (  { {\mu v - \bar x+x} \over {\sigma \sqrt{v}} } ) \biggl] ,
\end{split}
\end{equation}
where
\begin{align*}
&P_{n,k}=\sum _{i=k} ^{n-1} {n-k-1 \choose i-k} { n \choose i} ( { {\eta_+} \over {\eta_+ +\eta_ -} }  )^{i-k} ( { {\eta_-} \over {\eta_+ +\eta_ -} }  )^{n-i} p^i (1-p)^{n-i},\,\, 1 \le k \le n-1, ~P_{n,n}=p^n,\\
&Q_{n,k}=\sum _{i=k} ^{n-1} {n-k-1 \choose i-k} { n \choose i} ( { {\eta_+} \over {\eta_+ +\eta_ -} }  )^{n-i} ( { {\eta_-} \over {\eta_+ +\eta_ -} }  )^{i-k} p^{n-i} (1-p)^{i},\,\, 1 \le k \le n-1,~Q_{n,n}=(1-p)^n,
\end{align*}
\begin{align*}
 I_n(c;d,b,\delta) &=
   \begin{cases} - { e^{dc} \over d } \sum _{i=0} ^{n}  ( {b \over d} )^{n-i} Hh_i(bc-\delta) + ( {b \over d})^{n+1}  { {\sqrt{2\pi}} \over b } e^{  { {d \delta} \over b } + { {\sigma ^2} \over {2b^2} }   } \Phi (  -bc+\delta + { d \over b }  )       & \text{if }  b>0, \, d \neq 0, \\
   - { e^{dc} \over d } \sum _{i=0} ^{n}  ( {b \over d} )^{n-i} Hh_i(bc-\delta) + ( {b \over d})^{n+1}  { {\sqrt{2\pi}} \over b } e^{  { {d \delta} \over b } + { {\sigma ^2} \over {2b^2} }   } \Phi (  bc-\delta - { d \over b }  )        & \text{if }  b<0, \, d < 0,
  \end{cases}
\end{align*}
with $Hh_n(x)={{(n!)^{-1}} } \int _{x} ^{ \infty} (t-x)^n e^{-t^2/2} dt$, $\pi_n = e^{-\alpha v} (\alpha v)^n / n!$ and $v=\tilde T -t$.
}\\

\bibliographystyle{apa}
\linespread{0}
\begin{small}
\bibliography{mybib_2011}
\end{small}
\end{document}